\documentclass[aps,prd,twocolumn,amssymb,eqsecnum,showpacs,showkeyes,secnumarabic,graphics,floatfix,nofootinbib,tightenlines,longbibliography]{revtex4-1}
\usepackage{graphicx}
\usepackage{bm}
\usepackage{cancel}
\usepackage{epsfig}
\usepackage{dcolumn}
\usepackage{amsmath}
\usepackage{hhline}
\usepackage{enumerate}
\usepackage[utf8]{inputenc}
\usepackage[dvipsnames]{xcolor}
\usepackage[breaklinks=true,colorlinks=true,
linkcolor=blue,urlcolor=Blue,citecolor=MidnightBlue,% PDF VIEW
bookmarks=true,bookmarksopenlevel=2]{hyperref}
\usepackage{bm}
\usepackage{caption}
\usepackage{amsmath}
\usepackage {amssymb}
\usepackage{subcaption}

\def \beq  {\begin{equation}}
\def \eeq  {\end{equation}}
\def \ber  {\begin{eqnarray}}
\def \eer  {\end{eqnarray}}
\def \Geff {G_{\rm eff}}

\def \om    {\Omega}

\def \Geff {G_{\rm eff}}

\def \om0m {\Omega_{0\rm m}}

\begin{document}
\newcommand{\newc}{\newcommand}

\newc{\be}{\begin{equation}}
\newc{\ee}{\end{equation}}
\newc{\ba}{\begin{eqnarray}}
\newc{\ea}{\end{eqnarray}}
\newc{\bea}{\begin{eqnarray*}}
\newc{\eea}{\end{eqnarray*}}
\newc{\D}{\partial}
\newc{\ie}{{\it i.e.} }
\newc{\eg}{{\it e.g.} }
\newc{\etc}{{\it etc.} }
\newc{\etal}{{\it et al.}}
\newc{\lcdm}{$\Lambda$CDM }
\newc{\lcdmnospace}{$\Lambda$CDM}
\newc{\wcdm}{$w$CDM }
\newc{\plcdm}{Planck15/$\Lambda$CDM }
\newc{\plcdmnospace}{Planck15/$\Lambda$CDM}
\newc{\wlcdm}{WMAP7/$\Lambda$CDM }
\newc{\wlcdmnospace}{WMAP7/$\Lambda$CDM}
\newcommand{\fs}{{\rm{\it f\sigma}}_8}

\newcommand{\nn}{\nonumber}
\newc{\ra}{\Rightarrow}
\title{Hints of a Local Matter Underdensity or Modified Gravity in the Low $z$ Pantheon data}

\author{L. Kazantzidis}\email{l.kazantzidis@uoi.gr}
\affiliation{Department of Physics, University of Ioannina, GR-45110, Ioannina, Greece}
\author{L. Perivolaropoulos}\email{leandros@uoi.gr}
\affiliation{Department of Physics, University of Ioannina, GR-45110, Ioannina, Greece}

\date{\today}

\begin{abstract}
A redshift tomography of the Pantheon type Ia supernovae (SnIa) data focusing on the best fit value of the absolute magnitude $M$ and/or Hubble constant $H_0$ in the context of \lcdm indicates a local variation ($z\lesssim 0.2$) at $2\sigma$ level, with respect to the best fit of the full dataset. If this variation is not due to a statistical fluctuation, it can be interpreted as a locally higher value of $H_0$ by about $2\%$, corresponding to a local matter underdensity $\delta \rho_0/\rho_0 \simeq -0.10 \pm 0.04$. It can also be interpreted as a time variation of Newton's constant which implies an evolving Chandrasekhar mass and thus an evolving absolute luminosity $L$ and absolute magnitude $M$ of low $z$ SnIa. The local void scenario would predict a degree of anisotropy in the best fit value of $H_0$ since it is unlikely that we are located at the center of a local spherical underdensity. Using a hemisphere comparison method, we find an anisotropy level that is consistent with simulated isotropic Pantheon-like datasets. We show however, that the anisotropic sky distribution of the Pantheon SnIa data induces a preferred range of directions even in simulated Pantheon data obtained in the context of isotropic \lcdmnospace. We thus construct a more isotropically distributed subset of the Pantheon SnIa and show that the preferred range of directions disappears. Using this more isotropically distributed subset we again find no evidence for statistically significant anisotropy using either the hemisphere comparison method or the dipole fit method. In the context of the modified gravity scenario, we allow for an evolving normalized Newton's constant consistent with General Relativity (GR) at early and late times $\mu(z) = G_{\rm{eff}}(z,g_a)/G_{\rm{N}}=1+g_a z^2/(1+z)^2-g_a z^4/(1+z)^4$ and fit for the parameter $g_a$ assuming $L\sim G_{\rm{eff}}^b$. For $b=-3/2$ indicated by some previous studies we find $g_a=-0.47 \pm 0.36$ which is more than $1.5\sigma$ away from the GR value of $g_a=0$. This weak hint for weaker gravity at low $z$ coming from SnIa is consistent with similar evidence from growth and weak lensing cosmological data.
\end{abstract}
\maketitle

\section{Introduction}
\label{sec:Introduction}
Since the discovery of the accelerating expansion of the Universe \cite{Riess:1998cb,Perlmutter:1998np}, the \lcdm model based on the existence  of a cosmological constant \cite{Carroll:2000fy} has been particularly simple and consistent with most cosmological observations  including Cosmic Microwave Background (CMB) perturbations \cite{Hinshaw:2012aka,Ade:2015xua,Aghanim:2018eyx}, Type Ia supernovae (SnIa) standard candles \cite{Betoule:2014frx,Scolnic:2017caz} and Cosmic Chronometer probes of the expansion rate $H(z)$ \cite{Stern:2009ep,Moresco:2012jh,Gomez-Valent:2018hwc}, Baryon Acoustic Oscillations (BAO) standard ruler probes of $H(z)$ \cite{Alam:2016hwk,Aubourg:2014yra}, Large Scale Matter perturbations observed through Redshift Space Distortions (RSD) \cite{Basilakos:2016nyg,Aghanim:2018eyx,Quelle:2019vam}, Weak Lensing (WL) \cite{Baxter:2016ziy,Efstathiou:2017rgv}, Cluster Count data \cite{Rapetti:2008rm,Rozo:2009jj,Ade:2015fva} etc. Despite of its overall success and simplicity, the \lcdm model faces challenges on both theoretical and observational grounds. Theoretical challenges of \lcdm include the fine tuning \cite{Weinberg:1988cp,Sahni:2002kh} and  coincidence \cite{steinhardt,Velten:2014nra} problems, leading to a large variety of alternative theories attempting to solve these problems, see \eg Refs. \cite{Weinberg:1988cp,ArmendarizPicon:2000dh,Zimdahl:2000zm,Mangano:2002gg,Zhang:2005rj,Moffat:2005ii,Grande:2006nn,Amendola:2006dg,CalderaCabral:2008bx,Benisty:2018qed,Anagnostopoulos:2019myt,Hanimeli:2019wrt}. Observationally, there have been indications that different cosmological observations favour different values for the basic parameters of the model (at a level of $2\sigma$ or more) \cite{Risaliti:2018reu,Arjona:2019fwb,DiValentino:2019qzk,Handley:2019tkm,Arjona:2020kco} indicating that new degrees of freedom may be required to make the model simultaneously consistent with all these observations. These ``tensions" of \lcdm include the following:
\begin{itemize}
\item
{\bf The $H_0$ problem $>4\sigma$:} CMB and BAO cosmological measurements using the last scattering sound horizon as a standard ruler and assuming a \lcdm background expansion, report $H_0=67.4 \pm 0.5\;km \, s^{-1}\, Mpc^{-1}$ \cite{Aghanim:2018eyx}, a best fit value which is about $9\%$ lower compared to the local measurement of $H_0$ coming from SnIa data, that publish $H_0=74.03 \pm 1.42\;km \, s^{-1}\, Mpc^{-1}$ \cite{Riess:2019cxk}. The discrepancy ranges from $4.4 \sigma$ to more than $5 \sigma$ \cite{Wong:2019kwg,Camarena:2019moy,Riess:2020sih} depending on the combination of local data considered.  A similar value of $H_0 \approx 72 \pm 2\;km \, s^{-1}\, Mpc^{-1}$ is reported by string lens systems and time delay measurements \cite{Liao:2019qoc,Liao:2020zko}. Nevertheless, independent measurements of cosmic chronometers (based on models of evolving galaxy star luminosity) report a best fit value of $H_0=67.06 \pm 1.68\;km \, s^{-1}\, Mpc^{-1}$ \cite{Gomez-Valent:2018hwc} favouring the CMB and BAO measurements. On the contrary, measurements of $H_0$ based on a combination of cosmological data including a calibration of the Tip of the Red Giant Branch which is applied on SnIa (instead of the Cepheid calibration method) \cite{Freedman:2019jwv},  quasars, time-delay measurements, cosmic chronometers as well as $\gamma$ ray bursts \cite{Dutta:2019pio,Yang:2019vgk,Krishnan:2020obg} report a value that is intermediate between the CMB BAO and local measurements.
\item
{\bf The growth tension $\simeq 3\sigma$:} The growth rate and magnitude of linear cosmological perturbations depend on the matter density parameter $\Omega_{0m}$  and on the amplitude of the primordial power spectrum which is measured through the parameter $\sigma_8$, the linear amplitude of matter fluctuations on scales $8 h^{-1} Mpc$.
Weak Lensing (WL) \cite{Hildebrandt:2016iqg,Kohlinger:2017sxk,Joudaki:2017zdt,Abbott:2017wau,Abbott:2018xao} and  Redshift Space Distortion (RSD) \cite{Macaulay:2013swa,Johnson:2015aaa,Tsujikawa:2015mga,Wang:2016lxa,Sola:2016zeg,Basilakos:2017rgc,Nesseris:2017vor,Kazantzidis:2018rnb,Perivolaropoulos:2019vkb,Kazantzidis:2019dvk} cosmological observations measuring directly the growth rate of cosmological perturbations (dynamical probes) indicate that the observed growth rate is weaker than expected in the context of \lcdm with parameters determined from the observed background expansion rate using geometric probes (SnIa, BAO and CMB standard ruler data). This discrepancy is expressed as a preference for lower values of the parameters $\Omega_{0m}$ and $\sigma_8$ by dynamical probes compared to the corresponding values favoured by the geometric probes. The level of the growth tension is at least about $2-3\sigma$ \cite{Macaulay:2013swa,Johnson:2015aaa,Tsujikawa:2015mga,Nesseris:2017vor,Kazantzidis:2018rnb,Basilakos:2017rgc} but it can vary up to about $5\sigma$ (when the $E_G$ statistic data are used \cite{Skara:2019usd}) depending on the model parametrization and the type of dataset considered. Notice however that if CMB constraints on the background \lcdm parameters are not taken into account while keeping only background constraints from SnIa, the tension of the best fit \lcdm model with the growth data in the context of GR decreases to a level below $2\sigma$  \cite{LHuillier:2017ani}. Similarly, the tension decreases if marginalized confidence contours are used \cite{Quelle:2019vam}.
\item
{\bf Low-z galaxy BAO vs high z Ly$\alpha$ BAO curiosity ($\simeq 2\sigma$):} There is a $ \sim 2\sigma$ tension \cite{Addison:2017fdm,Cuceu:2019for} between the value of $\Omega_{0m}$ favoured by Ly-$\alpha$ BAO measurements ($\Omega_{0m} \simeq 0.19 \pm 0.07$ for $z>2.4$), which favour lower values of $\Omega_{0m}$, and the values of $\Omega_{0m}$ favoured by galaxy BAO measurements ($\Omega_{0m} \simeq 0.37 \pm 0.07$ for $z<0.6$) that favour higher values of $\Omega_{0m}$.
\item
{\bf Low l – high l CMB power spectrum curiosity ($\simeq 2\sigma$):} There is a mismatch of the cold dark matter density parameter $\Omega_c h^2$ best fit that is derived using high ($l>1000$) and low multipoles ($l<1000$). This  tension is approximately at a $2.5 \sigma$ level \cite{Addison:2015wyg,Aghanim:2019ame} and is such that the low $l$ multipoles predict a lower value of $\Omega_c h^2$ than the high-$l$ multipoles. It is also described by the need to introduce the $A_L$ parameter \cite{Aghanim:2019ame} which multiplies the amplitude of the lensing potential power \cite{Aghanim:2019ame}. Thus, this tension is also described by the fact that the high-$l$ TT multipoles are observed to correspond to a higher $\phi \phi$ lensing potential (the high-$l$ secondary peaks of the TT CMB power spectrum are smoother than expected in the context of the best fit  \plcdm model parameters). The value of the best fit Hubble parameter is also about $2.5\sigma$ lower when obtained from the higher-$l$ multipoles ($H_0=64.1\pm 1.7km \; s^{-1} \; Mpc^{-1}$) compared to the corresponding best fit value obtained from the low-$l$ CMB spectrum multipoles ($H_0=69.7 \pm 1.7km \; s^{-1} \; Mpc^{-1}$) \cite{Addison:2015wyg}.
\end{itemize}
 
The strongest of the above tensions which has also been called a ``problem" due its persistence in time and its increasing statistical significance is the Hubble parameter tension. This is heavily based on the use of SnIa as standard candle probes of the cosmic expansion rate. SnIa have been extensively used as standard candles to probe the expansion rate (Hubble parameter) $H(z)$ of the late Universe ($z<2$). The theoretically predicted apparent magnitude $m_{th}(z)$ of SnIa is connected with the Hubble free luminosity distance $D_L(z)\equiv H_0 d_L(z)/c$ as
\be
m_{th}(z)=M +5 log_{10}\left[D_L(z)\right] + 5 log_{10}\left(\frac{c/H_0}{1Mpc}\right)+25
\label{mthz}
\ee
where $M$ is the colour and stretch corrected absolute magnitude of SnIa (assumed constant) and $d_L(z)$ is the luminosity distance of each SnIa which in a flat Universe is 
\be 
d_L(z)=c (1+z) \int_0^z\frac{dz'}{H(z')}
 \label{dlz} 
 \ee
Using Eq. (\ref{mthz}), measurements of the SnIa apparent magnitude  $m(z)$ at various redshifts can be used to determine the present day Hubble parameter $H_0$ as well as its redshift dependence through Eq. (\ref{dlz}). For the determination of $H_0$, Riess et. al \cite{Riess:2009pu} used local distance ladder measurements (Cepheid calibrations at $z\simeq 0.01$) to measure directly $M$ and then a kinematic local expansion of $D_L(z)$ as
\be
D_L(z)=z \left[ 1+\frac{1}{2} (1-q_0)z-\frac{1}{6}(1-q_0-3q_0^2+j_0)z^2 + ... \right]
\label{ddlz}
\ee
to fit for the parameters $H_0$, $q_0$, $j_0$ \cite{Visser:2003vq} using low $z$ SnIa $(z\lesssim 0.2)$. 

For the determination of cosmological parameters in $H(z)$, higher $z$ SnIa are used and the degenerate parameters $M$, $H_0$ are usually marginalized as nuisance parameters \cite{Conley:2011ku,Betoule:2014frx,Scolnic:2017caz}. For example, in the context of \lcdm with 
\be
H^2(z)=H_0^2 \left[\Omega_{0m} (1+z)^3 +(1-\Omega_{0m})\right]
\label{hzlcdm}
\ee
Eq. (\ref{mthz}) is used for the construction and minimization of
${\bar \chi}^2(\Omega_{0m})\equiv\int d{\cal M}\;  \chi^2({\cal M},\Omega_{0m})$ where the degenerate combination
\be
{\cal M} \equiv M +5 log_{10}\left[\frac{c/H_0}{1Mpc}\right]+25= M -5 log_{10}(h)+42.38
\label{calmdef}
\ee
($H_0=100h \;km \, s^{-1}\, Mpc^{-1}$) has been marginalized. 

The marginalization of the parameter $\cal M$ however can lead to loss of useful physical information related to possible spatial variations of $H_0$ and/or time variations of the absolute magnitude $M$. For example, a value of $\cal M$ that evolves with redshift in a way that leads to low $\cal M$ values at low $z$ could imply either higher local values of $H_0$ due to a local matter underdensity or lower values of the absolute magnitude $M$ at recent cosmological times due to \eg a time variation of Newton's constant. 

The former case is in agreement with a few independent groups that have found evidence for a local matter underdensity on scales $100-300 h^{-1} Mpc$ \cite{Shanks:2018rka,Shanks:2019inu} with $\delta \rho_0/\rho_0$ in the range between $-0.1$ and $-0.3$ using either SnIa \cite{Lukovic:2019ryg} in the context of a Lemaitre-Tolman Bondi (LTB) metric \cite{Lemaitre:1993,Tolman:1934za,Bondi:1947fta} ($\delta \rho_0/\rho_0 \simeq -0.15$) or galaxy survey catalogues \cite{Boehringer:2019xmx} to construct luminosity density samples in the redshift range of $0.01<z<0.2$ ($\delta \rho_0/\rho_0\simeq -0.3$). A local matter underdensity of about $15\%$ corresponds to a local variation (increase) of $H_0$ by about $2\%$ which is in the right direction but not large enough to explain the $H_0$ tension which would require a local increase of $H_0$ by about $9\%$ compared to its mean value in the Universe, \ie a much deeper underdensity than the one implied by SnIa data.

If the later case is realized in Nature, the evolution of the absolute magnitude $M$ of SnIa (or equivalently the absolute luminosity $L\sim 10^{-2 M/5}$) could be used as a probe of the evolution of fundamental constants like the fine structure $\alpha$ or the Newton's constant $\Geff$. In the physical context of an evolving $\Geff$, previous studies \cite{GarciaBerro:1999bq,Gaztanaga:2001fh} assumed that the amount of $^{56}Ni$ that is produced in a SnIa and determines the absolute luminosity $L$, is proportional to the Chandrasekhar mass $m_{ch}\sim G_{\rm{eff}}^{-3/2}$ which implies that $L$ will increase as $G_{\rm{eff}}$ decreases.  In contrast, other more recent studies \cite{Wright:2017rsu} using a semi-analytical model to obtain SnIa light curves in the context of modified gravity, have indicated that $L$ will increase as $G_{\rm{eff}}$ increases. Assuming a power law dependence $L(z)\sim G_{\rm{eff}}(z)^b$ and fixing the value of $b$, any detected redshift dependence of SnIa absolute luminosity (or equivalently absolute magnitude) can be translated into a redshift dependence of $G_{\rm{eff}}$.

Therefore, a possible detection of redshift dependence of the parameter $\cal M$ could imply either a local underdensity and spatially varying $H_0$ or a redshift dependent $M$ and thus, possibly, an evolving $\Geff$. Since $M$ and $H_0$ are degenerate parameters within $\cal M$ in Eq. (\ref{calmdef}), the two scenarios can not be distinguished using only the redshift dependence of the SnIa apparent magnitudes. However, the local matter underdensity scenario with an off-center observer generically predicts a level of anisotropy in the best fit value of the parameter $\cal M$ which would emerge due to the anisotropy of $H_0$ which is expected for an off center observer in a region of matter underdensity. Such an anisotropy could also manifest itself as an anisotropy of cosmological parameters entering $H(z)$ like the matter density parameter $\Omega_{0m}$. Despite intense efforts to identify such anisotropy in the latest SnIa data (the Pantheon compilation \cite{Sun:2018cha,Andrade:2018eta,Deng:2018jrp,Zhao:2019azy,Chang:2019utc} and the joint light-curve analysis of the SDSS-II and SNLS supernova samples data (JLA) \cite{Lin:2015rza,Deng:2018yhb,Sun:2018epo}) no such anisotropy has been identified at a statistically significant level \cite{Sun:2018cha,Andrade:2018eta,Deng:2018jrp,Zhao:2019azy,Chang:2019utc,Lin:2015rza,Deng:2018yhb,Sun:2018epo}. However, we stress that none of these analyses has attempted to identify anisotropy signals using the parameter $\cal M$ (or equivalently the parameters $H_0$ and/or $M$). In the present analysis we aim to fill this gap in the literature\footnote{Notice however that in the case of the DF method, the parameter $\cal{M}$ was taken into account in Refs. \cite{Zhao:2019azy,Chang:2019utc}}.

The main questions addressed in the present analysis include the following:
\begin{itemize}
\item
What is the level of statistical significance for an evolving with redshift parameter $\cal M$ in the context of the Pantheon SnIa dataset?
\item
Are there any hints for anisotropy for the parameter $\cal M$ (or equivalently the parameters $M$ and/or $H_0$) in the context of the Pantheon SnIa dataset? Such an anisotropy would favour the scenario of a local matter underdensity rather than evolving absolute magnitude $M$. What is the optimal method for detecting such a possible anisotropy?
\item
If any hint for evolving $\cal M$ is interpreted as a hint for evolving $M$ and evolving Newton constant what is the best fit of the evolving $\mu(z) \equiv G_{\rm{eff}}(z)/G_{\rm{N}}$ (where $G_{\rm{N}}$ is the value of Newton's constant measured on solar system scales) and does it correspond to weakening gravity at low $z$ as the growth and weak lensing cosmological data seem to indicate?
\end{itemize}
 
The structure of this paper is the following: In the next section we use various subsets of the Pantheon dataset, to obtain the possible redshift dependence of the best fit parameters $\cal M$ and $\Omega_{0m}$ in the context of \lcdmnospace. In section \ref{sec:voidmod} we use the hemisphere comparison (HC) method and the dipole fiting (DF) method to search for possible statistically significant directional dependence of the best fit parameter $\cal M$ (or equivalently the parameter $H_0$ with fixed $M$). Such an anisotropy would be generically anticipated in the context of a cosmological off-center observer in a local matter underdensity. In section \ref{sec:modgrav} we make the assumption that any variation of the parameter $\cal M$ is due to a variation of $M$ induced by a varying $\mu(z)$ and in the context of a physically motivated single parameter parametrization of $\mu(z)$ we identify the best fit parameter value and corresponding strength of gravity at low $z$ ($z\lesssim 0.2$) compared to the corresponding value at higher $z$. Finally, in section \ref{sec:concl}, we summarize, discuss the possible physical implications of our results and identify the possible extensions of this work. 

\section{Searching for a redshift dependence of $\cal M$}
\label{sec:maxlikelihood}
\begin{figure*}[ht!]
\centering
\includegraphics[width = 1.0 \textwidth]{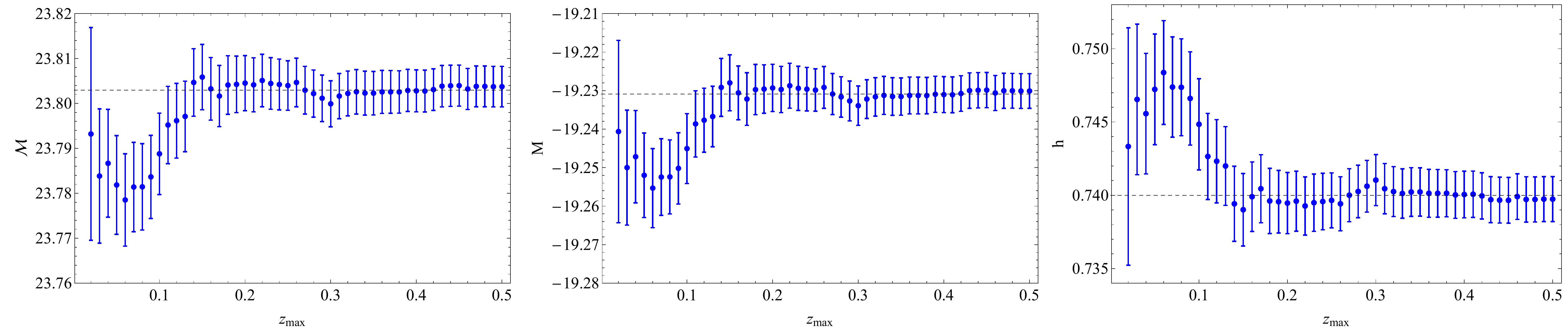}
\caption{The evolution of the best fit values (blue dots) of ${\cal{M}}$ (left panel), $M$ (middle panel) and $h$ (right panel) along with its $1\sigma$ error for various cutoff values $z_{max}$. The dashed lines correspond to the best fit values indicated by the full dataset.}
\label{fig:cutoff}
\end{figure*} 

The Pantheon dataset \cite{Scolnic:2017caz} is the largest compilation to date that incorporates data from six different probes giving a total of 1048 SnIa datapoints covering the redshift range $0.01<z <2.3$. The publicly available data include the name of each SnIa, the redshifts in the CMB and heliocentric frames as well as the observed corrected apparent magnitude $m_{obs}$ along with the corresponding error $\sigma_{m_{obs}}$. The $m_{obs}$ of each SnIa is reported after applying  color and stretch corrections as well as corrections due to biases from simulations of the SnIa. In the context of a maximum likelihood analysis \cite{Arjona:2018jhh} Eqs. \eqref{mthz}, \eqref{dlz} and \eqref{calmdef} are used to construct the appropriate $\chi^2$ function as 
\be 
\chi^2 ({\cal M},\Omega_{0m})=V^i_{Panth.} \, C_{ij}^{-1} \, V^j_{Panth.} \label{chifunc}
\ee 
where $V^i_{Panth.}\equiv m_{obs}(z_i)-m_{th}(z)$ and $C_{ij}$ is the covariance matrix which is given as $C_{ij}=\bar{D}_{ij}+\bar{C}_{sys}$, where $\bar{D}_{ij}$ is the diagonal matrix
\be 
\bar{D}_{ij}=\left(
         \begin{array}{cccc}
           \sigma_{m_{obs},1}^2 & 0 & 0 & \cdots \\
           0 & \sigma_{m_{obs},2}^2 & 0& \cdots \\
           0 & 0 & \cdots &   \sigma_{m_{obs},N}^2 \\
         \end{array}
       \right) \label{eq:totalcij}
\ee
and $\bar{C}_{sys}$ is a non-diagonal matrix associated with the systematic uncertainties that emerge from the bias corrections method (see Ref. \cite{Scolnic:2017caz} for more details). In what follows we consider statistical uncertainties only. This approach makes the analysis much simpler due to the diagonal nature of the covariance matrix  but leads to somewhat lower uncertainties of the derived best fit parameters. In Appendix \ref{sec:Appendix_A} we have included a short analysis which takes into account systematic uncertainties. This analysis indicates that systematic effects tend to somewhat increase the uncertainties of the best fit parameter values. The main features and conclusions, however, of the analysis presented below remain valid. 

As discussed in the Introduction there is a power law dependence of the absolute luminosity $L$ on $\Geff$ leading to a simple power law relation between $M$ and $\Geff$. For $L\sim G_{\rm{eff}}^b$ this equation is of the form 
\be 
M-M_0=-\frac{5 \, b}{2} \, log_{10} \left(\mu \right) \label{absmagngeffgen}
\ee
where $M_0$ corresponds to a reference local value of the absolute magnitude. Then, Eq. \eqref{mthz} takes the following form
\be 
m_{th}(z)={\cal{M}}+5 log_{10} \left[D_L(z)\right]-\frac{5b}{2} \, log_{10} \left(\mu \right) \label{appmagthmodgen}
\ee
where ${\cal M}$ is given in Eq. \eqref{calmdef} with $M$ replaced by $M_0$.

Most previous studies used $b=-3/2$ \cite{GarciaBerro:1999bq,Gaztanaga:2001fh} based on the assumption that $L\sim m_{ch}\sim G_{\rm{eff}}^{-3/2}$. However, as mentioned in the Introduction, a more detailed analysis has been performed in \cite{Wright:2017rsu}, where the authors studied the effects of modified theories of gravity to the absolute magnitude $M$ of the SnIa. In particular, in the  semi-analytic model that was used, extra parameters such as  the initial nickel mass in the ejecta, the initial radius of shock breakout, the scale velocity, the effective opacity as well as  total ejecta mass were included. Then, the generated light curves were standardised by rescaling the shape to match a template width and the numerical dependence of the standardised intrinsic absolute luminosity $L$ on $\Geff$ was identified. Using this semi-analytical method a new power law relation between $L$ and $\Geff$ was derived with $b>0$ (see the left panel of Fig. 7 of \cite{Wright:2017rsu}).

A marginalisation is usually implemented over $\cal M$ in most analyses of the SnIa data (\eg  \cite{Conley:2011ku}). This approach however, may lead to loss of useful information regarding possible redshift dependence of $H_0$ and/or $M$ and $\Geff$. Thus, we choose to keep this parameter and fit it along with the cosmological parameter $\Omega_{0m}$. Fixing  the background to that of a \lcdmnospace, we implement the maximum likelihood method \cite{Arjona:2018jhh} to obtain the best fit values for $\Omega_{0m}$ and $\cal{M}$ as ${\cal{M}}=23.803 \pm 0.007$ and $\Omega_{0m}=0.285 \pm 0.012$ for the full Pantheon dataset in agreement with previous studies \cite{Scolnic:2017caz,Zhao:2019azy}.

\begin{figure*}
\centering
\includegraphics[width = 1.0\textwidth]{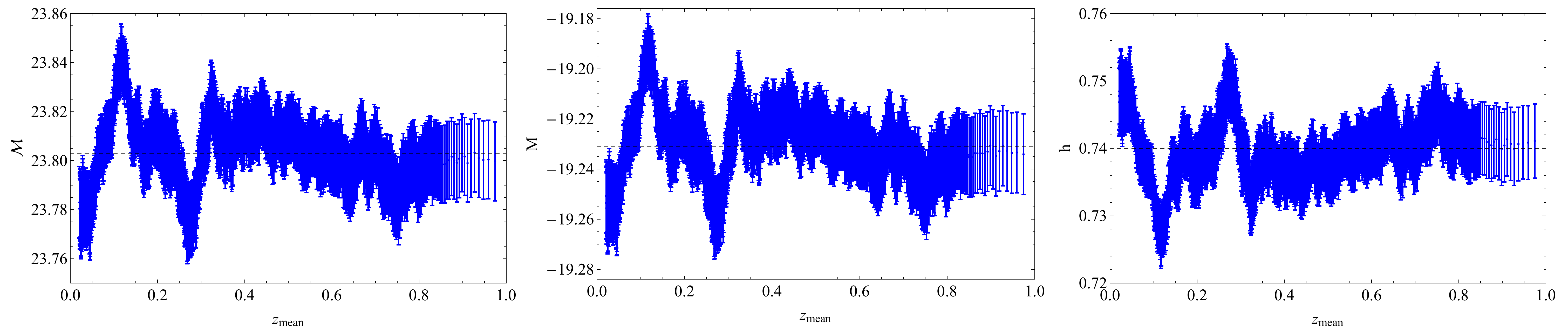}
\caption{The evolution of the best fit values (blue dots) of ${\cal{M}}$ (left panel), $M$ (middle panel) and $h$ (right panel) along with its $1\sigma$ error for 100 point subsamples vs the mean redshift $z_{mean}$. The dashed lines correspond to the best fit values indicated by the full dataset.}
\label{fig:subsamp}
\end{figure*} 

In the context of a redshift independent ${\cal{M}}$ and a \lcdm background, any subset of the Pantheon dataset should provide  best fit parameter values for $\Omega_{0m}$ and $\cal{M}$ consistent with the corresponding best fit values of the full dataset. In order to test this conjecture, we fix  $\Omega_{0m}$  to its best fit value indicated by the full dataset and consider subsets of the full dataset in redshift ranges $z\in [z_{min},z_{max}]$ where $z_{min}=0.02$ (fixed) and $z_{max}\geq 0.03$ (increasing for each point in steps of $\Delta z_{max}=0.01)$ is a cutoff redshift chosen so that the subsamples have acceptable statistics (the first and smallest subsample with $z_{max}=0.03$ has 46 datapoints). Using each subsample, we implement the maximum likelihood method to find the best fit $\cal M$ values along with their $1\sigma$ errors shown in Fig. \ref{fig:cutoff} (left panel). For the best fit values of $M$ of each subsample (Fig. \ref{fig:cutoff}  middle panel)  we fix $h=0.74$, \ie to the value specified in \cite{Riess:2019cxk}, while for the best fit values of $h$ (right panel of Fig. \ref{fig:cutoff}) we fix $M$ using the best fit value of $\cal M$ indicated by the full dataset and $h=0.74$.

Clearly, at low redshifts and in particular in the redshift range $z_{max} \in [0.02,0.15]$ there is a tension of about $2 \sigma$ or more between the best fit value of $\cal{M}$ of each subsample and the best fit value indicated by the full dataset. This difference may imply a lower value of $M$ (middle panel of Fig. \ref{fig:cutoff}) or equivalently a higher value of $h$ (right panel of Fig. \ref{fig:cutoff}) in the same redshift range. For $z_{max}>0.15$ the best fit values in each subsample are consistent with the values indicated by the full dataset (dashed lines in Fig. \ref{fig:cutoff}) within $1 \sigma$ level.

\begin{figure*}
\centering
\includegraphics[width =0.9\textwidth]{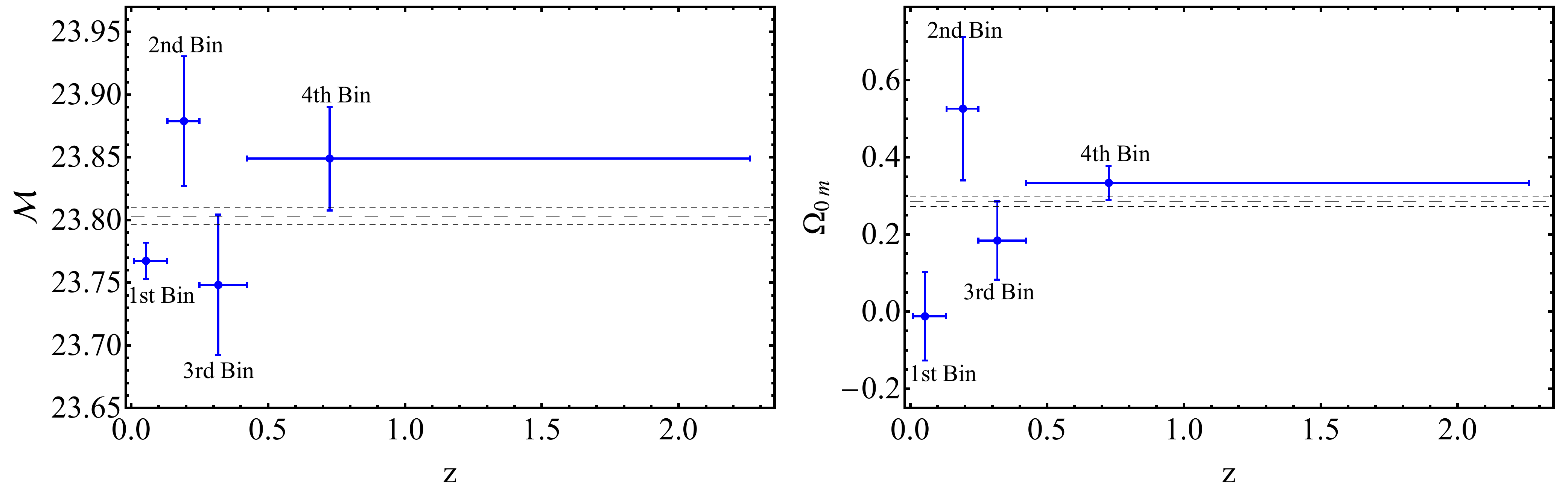}
\caption{The best fit values of $\cal{M}$ (left panel) and $\Omega_{0m}$ (right panel) as well as the $1 \sigma$ errors for the four bins. The horizontal axis corresponds to the redshift range of each bin.The dashed line describes the best fit value of the full dataset while the dot dashed lines its $1\sigma$ error. The corresponding plot, taking into account the systematic uncertainties is shown in the Appendix \ref{sec:Appendix_A} and shows a similar oscillating behaviour of the parameters with increased uncertainties (specifically for the lowest $z$ bin)}
\label{fig:crossplot}
\end{figure*} 

A similar behaviour is detected, if we rank the SnIa data from  lowest to highest redshifts. At first we select the first 100 datapoints and fixing the background to the best fit \lcdm $H(z)$, [Eq. \eqref{hzlcdm} with $\Omega_{0m}=0.285$], we find the best fit value of $\cal{M}$ along with its $1 \sigma$ error for the lowest redshift subsample. Then, we shift the 100 points subsample by one datapoint towards higher redshifts to produce the next point and continue until we cover the entire redshift range of the Pantheon dataset (Fig. \ref{fig:subsamp} - left panel). The redshift $z_{mean}$ shown in the horizontal axis, corresponds to the mean redshift value of each of the 100 point subsamples.

From Fig. \ref{fig:subsamp} we observe that for $z_{mean}<0.3$, the best fit value of $\cal{M}$ oscillates around the best fit value of the full dataset  at a level of about $1-2\sigma$ which may indicate a similar oscillating behaviour for $M$ (middle panel) and/or $h$ (right panel) in the same redshift range. In this case, the redshift range of the oscillation is larger than the redshift variation detected in Fig. \ref{fig:cutoff}, because as the cutoff redshift increases, so does the size of the corresponding subsample, leading to a cancellation of the oscillating effect in Fig. \ref{fig:cutoff}.

\begin{figure*}[ht!]
\centering
\includegraphics[width = 1.03\textwidth]{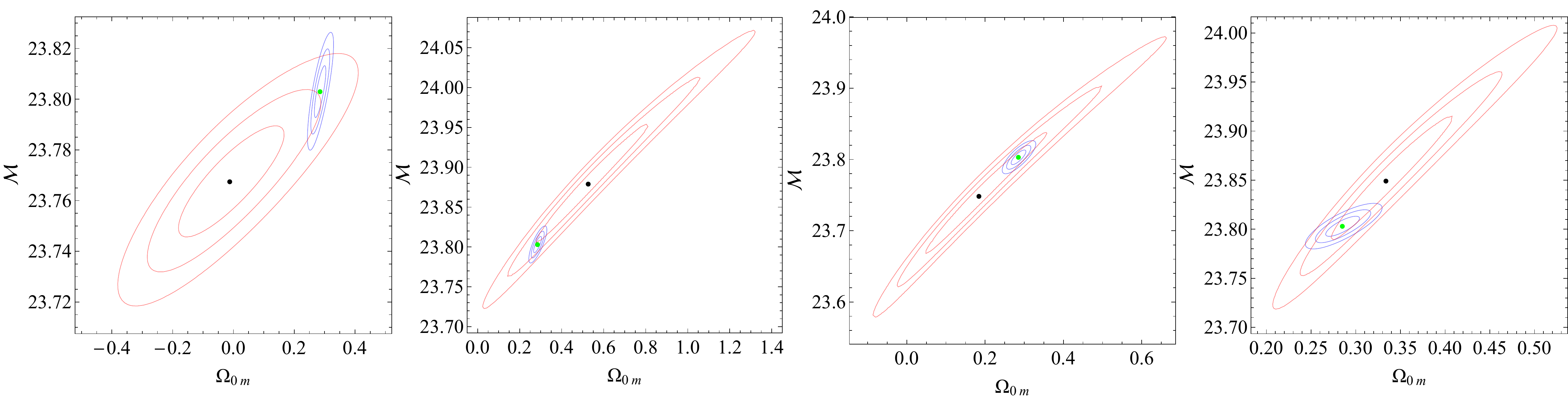}
\caption{The $1\sigma-3\sigma$ confidence contours in the parametric space $ \left(\Omega_{0m}-{\cal{M}} \right)$. The blue contours correspond to the $1\sigma-3\sigma$ full Pantheon dataset best fit, while the red contours describe the $1\sigma-3\sigma$ confidence contours of the four bins (from left to right). The black points represent the best fit of each bin, while the green dot represents the best fit value indicated by the full Pantheon dataset ($\Omega_{0m}=0.285$ and ${\cal{M}}=23.803$).}
\label{fig:contourtomfig}
\end{figure*}

\begin{figure*}[ht!]
\centering
\includegraphics[width = 1.04\textwidth]{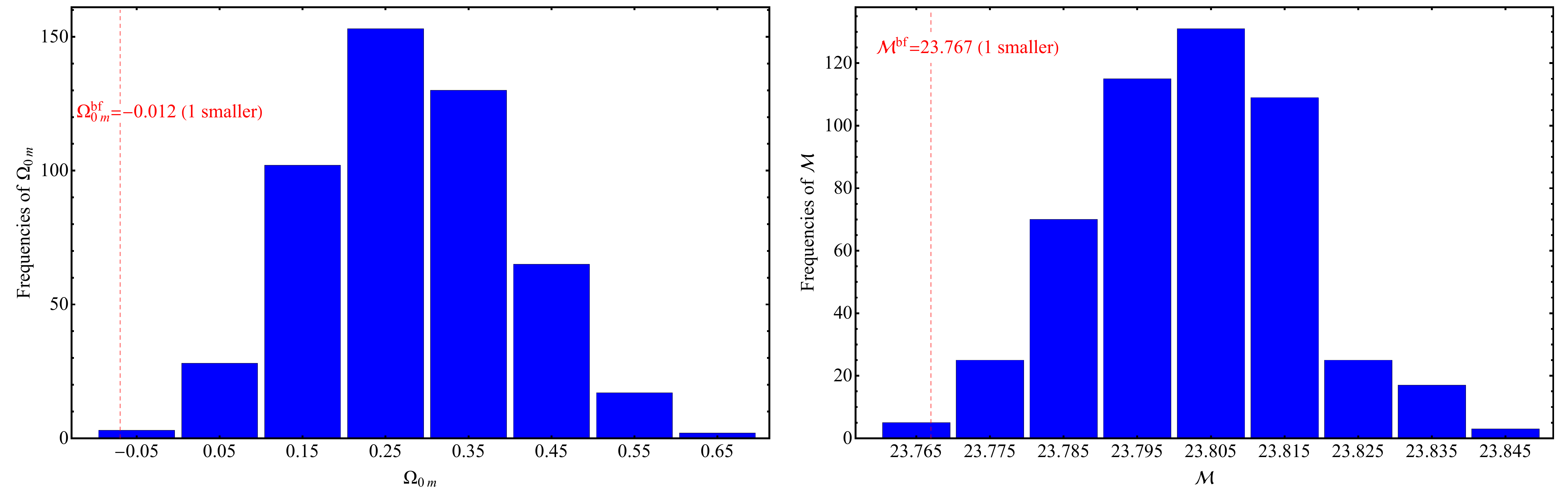}
\caption{The distributions of $\Omega_{0m}$ (left panel) and $\cal{M}$ (right panel) using $500$ Monte Carlo simulations of Pantheon-like datasets under the assumption of an underlying \lcdm model, in the redshift region $0.01<z<0.13$ (first bin). The red dashed lines correspond to the best fit values of the first bin.}
\label{fig:barchartfig}
\end{figure*}
In order to improve the statistics of the low $z$ subsamples and further investigate the observed tension at low $z$, we sort the Pantheon data from lowest to highest redshift  and divide them in four equal uncorrelated bins consisting of 262 datapoints. Then, we apply the maximum likelihood method in each bin separately, considering a \lcdm background and leaving the parameters $\cal{M}$ and $\Omega_{0m}$ to vary simultaneously. Minimizing Eq. \eqref{chifunc}, we derive the best fit values of $\Omega_{0m}$ and $\cal{M}$ as well as the corresponding $1 \sigma$ error for each bin as it is shown in Fig. \ref{fig:crossplot}. 

Clearly a similar oscillating behaviour for $\cal{M}$ is apparent as in Fig. \ref{fig:subsamp}. Furthermore,  the best fit values of $\Omega_{0m}$ and $\cal{M}$ derived from the lowest $z$ bin $(0.01<z<0.13)$ are more than $2 \sigma$ lower than the best fit values indicated by the full dataset, in agreement with Figs. \ref{fig:cutoff} and \ref{fig:subsamp}. This is also evident in Fig. \ref{fig:contourtomfig}, where the $1\sigma-3\sigma$ contours of the four bins are constructed in the parametric space $\left(\Omega_{0m}-{\cal{M}}\right)$. 

From Fig. \ref{fig:crossplot} we find the difference of $\cal{M}$ to be
\be 
 \Delta {\cal{M}} \equiv {\cal{M}}_{bf}-{\cal{M}}_{bin1} \approx 23.80-23.76 \approx 0.04 \pm 0.02
 \ee
where ${\cal{M}}_{bf}$ corresponds to the best fit value of $\cal{M}$ indicated by the full dataset and ${\cal{M}}_{bin1}$ corresponds to the best fit value of ${\cal{M}}$ derived from the lowest $z$ bin. In the context of a local matter underdensity, ${\cal{M}}_{bf}$ is the true global value of $\cal{M}$, while ${\cal{M}}_{bin1}$ corresponds to the 
value of $\cal{M}$ that is measured in the interior of the local underdensity. This difference can be associated with a variation of the local expansion rate $\delta H_0/H_0$ through
\be 
\left(\frac{\delta H_0}{H_0} \right) \approx 0.2 \, ln(10) \Delta {\cal{M}}=0.019 \pm 0.007  \label{eq:deltahoverh}
\ee

We can also estimate the propability that such low values of $\cal{M}$ and $\Omega_{0m}$ would occur in the context of the \lcdm standard model, using Monte Carlo simulations of Pantheon-like datasets under the assumption of an underlying \lcdm model with $\cal{M}$ and $\Omega_{0m}$ corresponding to the values indicated by the full dataset. In particular, we construct 500 simulated datasets with redshifts corresponding to the redshifts of the first bin and substitute the apparent magnitude of the real data [$m_{obs}(z_i)$] with simulated datapoints $m_{sim}(z_i)$ obtained from a random normal distribution with a mean value obtained from the best fit \lcdm value of the apparent magnitude $m_{th}$ [setting ${\cal{M}}=23.803$ and $\Omega_{0m}=0.285$ in Eq. \eqref{mthz}]. The standard deviation of the normal distribution is obtained from the $\sigma_{m_{obs}}$ of each datapoint respectively \cite{Antoniou:2010gw}. Then, we apply the maximum likelihood method and count how many of the simulated data give lower values for $\cal{M}$ and $\Omega_{0m}$ than the best fit values indicated by the real data of the first bin (red dashed lines). The results are plotted in Fig. \ref{fig:barchartfig}. Clearly, less than $1\% $ ($0.2 \%$ for either $\Omega_{0m}$ or $\cal{M}$) of the Monte Carlo data give smaller best fit values for $\cal{M}$ or $\Omega_{0m}$ than the actual best fit values of the first bin. Therefore, we confirm that this reduced value of $\cal{M}$ is a highly unlikely event in the context of an underlying physical \lcdm model.

The $2-3\sigma$ effect regarding the parameters $\Omega_{0m}$ and $\cal{M}$ observed at low $z$ have been also discussed in previous studies \cite{Colgain:2019pck,Lukovic:2019ryg}. In particular, for $\Omega_{0m}$  a similar behaviour was presented in \cite{Colgain:2019pck}, where the best fit values of $\Omega_{0m}$ and $H_0$ were studied, for various redshift cutoffs. Similar results for $\cal{M}$ were also presented in \cite{Lukovic:2019ryg}, where the authors divided the Pantheon dataset in three bins and calculated the best fit value of ${\cal{M}} -25$ in the context of a LTB model  with a cosmological constant, in an attempt to identify hints of a local underdensity using the Pantheon dataset.

This variation of $\cal{M}$ at low redshifts could be due to following:
\begin{itemize}
\item Statistical and or systematic fluctuations of the data around the true \lcdm model. The probability of this case can be estimated by constructing a large number of simulated Pantheon datasets under the assumption of a \lcdm underlying model corresponding to the best fit with a multivariate Gaussian distribution taking into account the full covariance matrix including both statistical and systematic errors. Our preliminary analysis along these lines (Fig. \ref{fig:barchartfig}) taking into account only statistical errors has indicated that this case is very unlikely (has a probability less than $1\%$). However, this probability is expected to increase if the simulated data are constructed taking into account also systematic errors and if the ``look elsewhere effect" is taken into account. Such an extension of our analysis is currently in progress.
\item A local underdensity dubbed ``Local Void" that fades away at large scales. Since $\cal{M}$ is lower than the best fit value indicated by the full dataset at low $z$, $h$ would be larger than the best fit value of the full dataset in the same redshift range. A generic way to explain this increase of $h$ would be if our neighbourhood is more underdense compared to the mean density of the Universe and as a result the measured value of $h$ is also affected at local scales. In the context of a ``Local Void" model, the value of $h$ increases by $2-3\%$ (see Fig. \ref{fig:subsamp} - right panel). Such a scenario would also predict an anisotropy for the best fit value of $\cal{M}$ in the sky.
\item A modified theory of gravity. Another possible explanation for the observed variation of $\cal{M}$ at low redshifts is a redshift dependence of $M$ which could be due to a time variation of Newton's constant in the context of a modified theory of gravity.
\end{itemize}
These two possibilities will be discussed in the next two sections.

\section{``Local Void" Scenario}
\label{sec:voidmod}
The idea that we live in an underdense region that fades away at large scales is not new. In fact, it has been proposed as an alternative theory to explain the accelerated expansion of the Universe without the presence of a cosmological constant (see \eg Refs. \cite{Alnes:2005rw,GarciaBellido:2008nz,Enqvist:2006cg,February:2009pv,Biswas:2010xm}). It has been shown however, the Gpc scale  and depth of the uderdensity required to explain the observed accelerating expansion is inconsistent with current observations \cite{Wu:2017fpr}. Nevertheless, over the past twenty years, some works using various galaxy survey catalogues (\eg the 2MASS survey \cite{Frith:2003tb,Frith:2005et}, the UKIDSS-Large Area Survey \cite{Keenan:2013mfa} as well as galaxies samples constructed from the 6dFGS, SDSS and GAMA surveys \cite{Whitbourn:2013mwa,Whitbourn:2016irk}) have found some evidence for the existence of a local underdensity that extends on scales $150-400 \, h^{-1} Mpc$ with depth $-0.4< \delta \rho_0/\rho_0<-0.05$.  Other works considering the Pantheon dataset \cite{Lukovic:2019ryg}  or a sample of 1653 X-ray galaxy clusters \cite{Boehringer:2019xmx} also stressed that a local underdensity on scales of $\approx 100 \, h^{-1} Mpc$ or $\approx 140 \, h^{-1} Mpc$ with $\delta \rho_0/\rho_0 \approx -0.11$ or $\delta \rho_0/\rho_0 \approx -0.20$ respectively, remains a viable possibility and can not be excluded by the data.

If this scenario is realized in Nature and we truly live in an underdense region, then the measured $H_0$ value at local scales would be larger than the true global value of $H_0$. This could lead to a lower value of $\cal{M}$ [Eq. \eqref{calmdef}] at local scales, explaining the results of the previous section. However, a slightly off-center observer in this underdense region would experience a preferred cosmological direction and an overall anisotropy. Therefore, in what follows we search for possible anisotropies regarding $\cal{M}$ using two different methods that are widely used in the literature. These are the Hemisphere Comparison (HC) \cite{Schwarz:2007wf,Antoniou:2010gw,Cai:2011xs,Chang:2014nca,Deng:2018jrp} and the Dipole Fitting (DF) \cite{Mariano:2012wx,Chang:2014nca,Lin:2015rza,Deng:2018jrp} method.
\subsection{Hemisphere Comparison (HC) Method}
The HC method was first proposed in Ref. \cite{Schwarz:2007wf} and implemented in the context of the Union2 dataset \cite{Amanullah:2010vv} in Ref. \cite{Antoniou:2010gw}. The basic steps of this method are the following:

\begin{figure*}
\centering
\includegraphics[width = 0.64\textwidth]{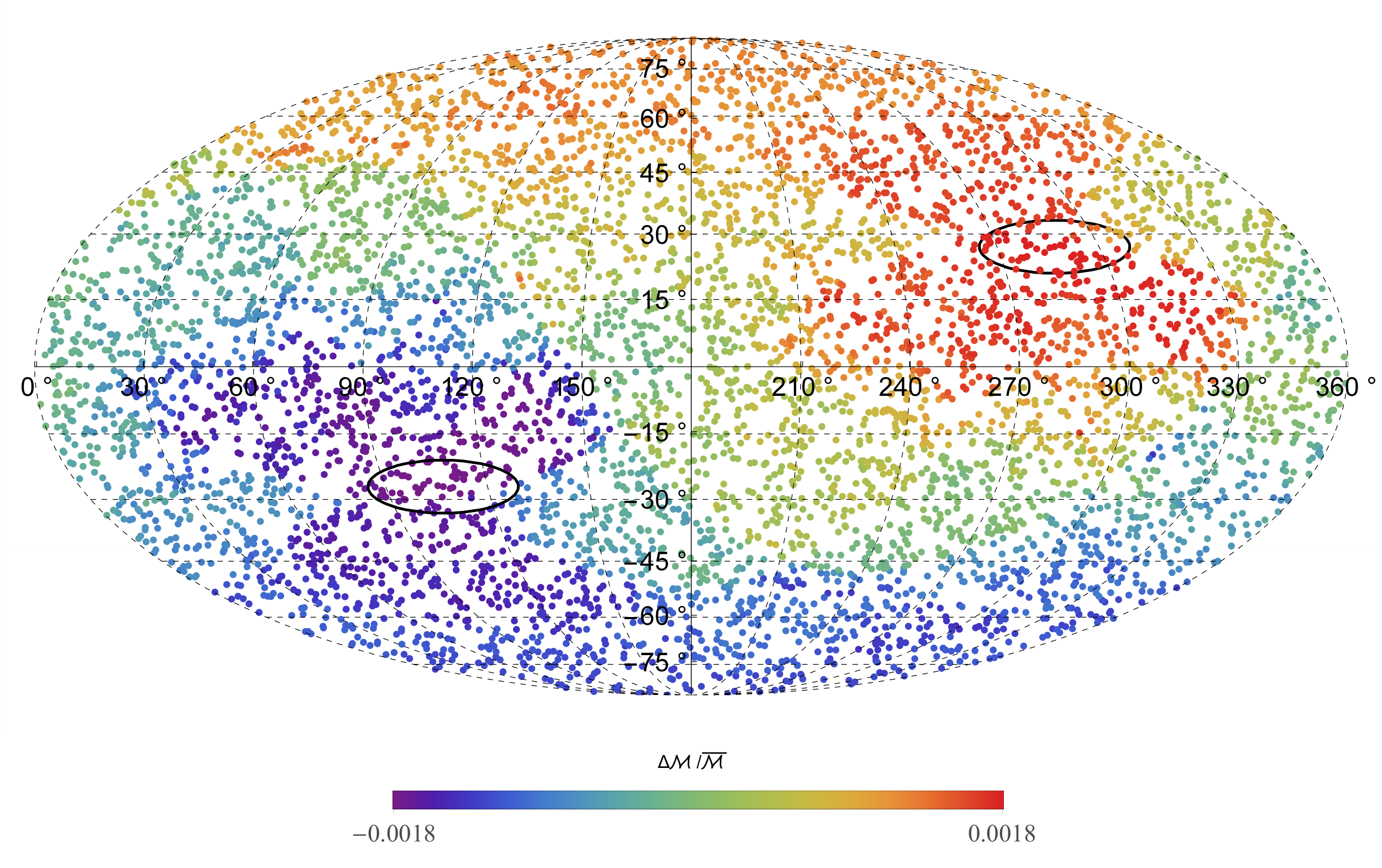}
\caption{The AL color map constructed with the HC method producing 3000 random directions. The red dots correspond to the pair of coordinates where the ratio $\Delta {\cal{M}}/\bar{{\cal{M}}}$ is maximum while the purple dots correspond to the  pair of coordinates where $\Delta {\cal{M}}/\bar{{\cal{M}}}$ is minimum. The black ellipses denote the $1 \sigma$ error region.}
\label{fig:maxdiral}
\end{figure*} 

\begin{itemize}
\item Consider a random direction of the following form
\be 
{\hat r}_{rndm}=(cos \phi \, \sqrt{1-cos^2 \theta}, \, sin\phi \, \sqrt{1-cos^2 \theta}, \, cos \theta)
\ee
where $\phi\in [0,2\pi]$ and $cos \theta\in [-1,1]$. These variables are randomly selected in these intervals with a uniform probability distribution.
\item Define two different hemispheres dubbed ``up hemisphere" and ``down hemisphere" and append the data of the dataset into the  hemisperes. The appended data of the ``up hemisphere" correspond to the subset where the product ${\hat r}_{rndm} \cdot {\hat r}_{data}$ is positive, while the appended data of the ``down hemisphere" correspond to the subset where the product ${\hat r}_{rndm} \cdot {\hat r}_{data}$ is negative. The unit vector ${\hat r}_{data}$ describes the direction of each SnIa in galactic coordinates.
\item Find the best fit value of $\cal{M}$ in the up (${\cal{M}}_{up}$) and down hemispheres (${\cal{M}}_{down}$) applying the maximum likelihood method for $\Omega_{0m}=0.285$, \ie setting  $\Omega_{0m}$ to the best fit value indicated by the full dataset. Using the obtained best fit values of $\cal{M}$, define the anisotropy level (AL) as \cite{Antoniou:2010gw}
\be 
\Delta {\cal{M}}/\bar{{\cal{M}}}  \equiv 2\frac{{\cal{M}}_{up}-{\cal{M}}_{down}}{{\cal{M}}_{up}+{\cal{M}}_{down}} \label{eq:al}
\ee
as well as the corresponding $1 \sigma$ error \cite{Antoniou:2010gw}
\be 
\sigma_{\Delta {\cal{M}}/\bar{{\cal{M}}}}=\frac{\sqrt{\sigma^2_{{\cal{M}}_{up}}+\sigma^2_{{\cal{M}}_{down}}}}{{\cal{M}}_{up}+{\cal{M}}_{down}}
\ee
\item Repeat this procedure for $N$ random directions ${\hat r}_{rndm}$ and find the maximum AL and the related direction. The number of random directions needs to be well above the number of datapoints in each hemisphere \cite{Antoniou:2010gw}, so for the Pantheon data we set $N=3000$.
\end{itemize}
Implementing the HC method in the Pantheon dataset as described above, we construct the AL color map of $\cal{M}$ as it is demonstrated in Fig. \ref{fig:maxdiral}. The magnitude of the maximum AL that is detected for the Pantheon data is 
\be 
\left(\Delta {\cal{M}}/\bar{{\cal{M}}} \right)_{max}=0.0018 \pm 0.0002
\ee
and the direction of the maximum anisotropy is in $(l,b)=(286.93^\circ \pm 18.52 ^\circ, 27.02 ^\circ \pm 6.50 ^\circ)$

\begin{figure*}
\centering
\includegraphics[width = 1.\textwidth]{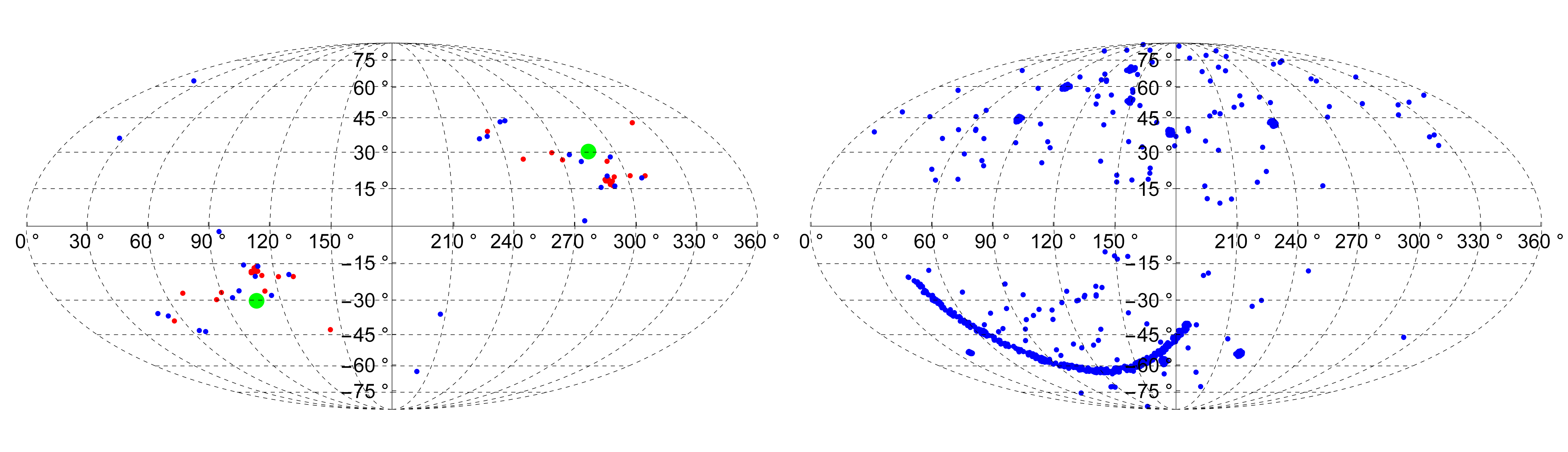}
\caption{\textit{Left Panel:} The 30 axes of extrema of AL constructed from the isotropic simulated Pantheon datasets axes using 3000 random hemisphere directions in each dataset. Notice that only two of the thirty maxima AL directions are in the lower left quadrisphere (southern hemisphere in the longitude range of $[0^\circ,180 ^\circ]$), inducing an artificial region of preferred directions in the observed anisotropy in the lower left/upper right quadrisphere. The green dot corresponds to the maximum anisotropy of the real data, while the blue (red) dots describe the simulated datasets which have smaller (larger) magnitudes of $\Delta {\cal{M}}/\bar{{\cal{M}}}$ than the real data. \textit{Right Panel:} The distribution of the full Pantheon data in galactic coordinates. Notice that the data are not uniformly distributed with strong preference of datapoint locations in the southern hemisphere in the longitude range $[0^\circ,180 ^\circ]$ (lower left hemisphere). }
\label{fig:distribfigreal}
\end{figure*}

\begin{figure*}
\centering
\includegraphics[width =1.0\textwidth]{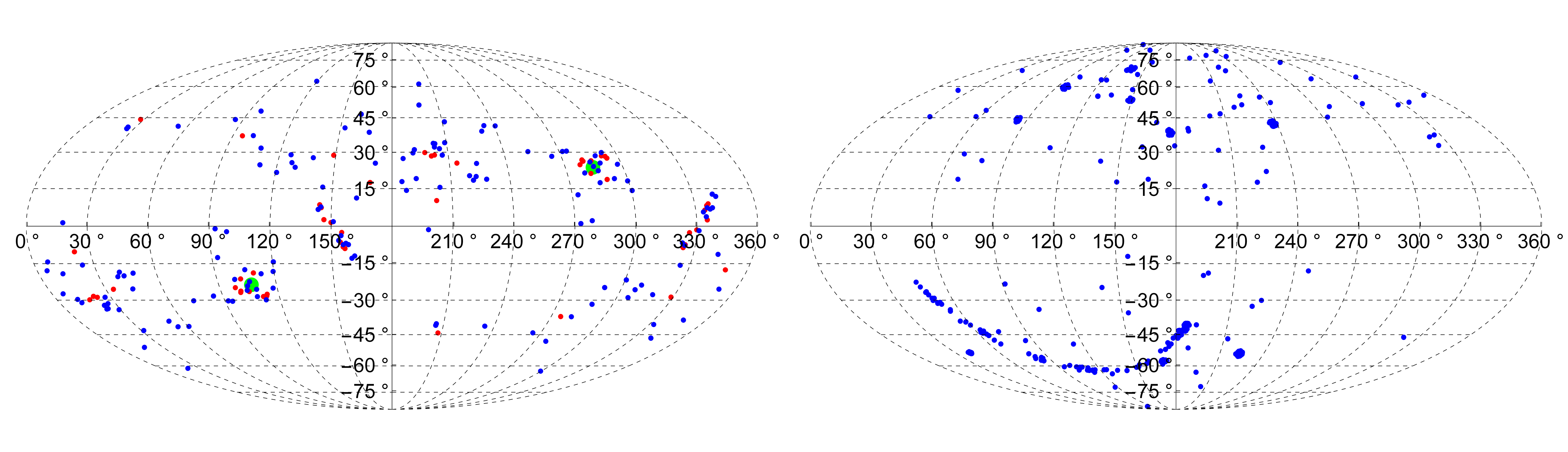}
\caption{\textit{Left Panel:} The 100 axes of extrema of AL using 1500 random directions for each isotropically distributed Pantheon subsample. The preferred direction disappears completely for the more isotropic distributed subset. The green dot corresponds to the maximum anisotropy of the real data, while the blue (red) dots describe the simulated datasets which have smaller (larger) magnitudes of $\Delta {\cal{M}}/\bar{{\cal{M}}}$ than the real data. \textit{Right Panel:} The distribution of the reduced isotropic subset in galactic coordinates.}
\label{fig:distribfigredreal}
\end{figure*} 

In order to check the consistency of the Pantheon SnIa data with statistical isotropy we compare the above extrema of AL of the real data with the corresponding extrema of AL derived in the context of simulated Pantheon data. The simulated Pantheon data are constructed under the assumption of statistical isotropy with a \lcdm background by keeping fixed the direction of each datapoint in the sky while randomly selecting the Pantheon apparent magnitudes from a gaussian distribution with the best fit \lcdm mean and standard deviation equal to the corresponding Pantheon datapoint $1 \sigma$ error. We thus construct 30 isotropic simulated ``Pantheon" datasets and for each dataset we use 3000 random directions to split it in two hemispheres and identify the corresponding extrema of AL using $\Delta {\cal{M}}/\bar{{\cal{M}}}$. These 30  axes of extrema of AL are shown in Fig. \ref{fig:distribfigreal} using galactic coordinates and showing two opposite points for each maximum AL direction  (left panel) along with the corresponding real Pantheon data sky directions. The maximum AL of $\Delta {\cal{M}}/\bar{{\cal{M}}}$ magnitude of 16 (red dots in the left panel of Fig. \ref{fig:distribfigreal}) out of the 30 simulated datasets was larger than the corresponding magnitude of the real Pantheon data. This indicates that there is no statistically significant $\Delta {\cal{M}}/\bar{{\cal{M}}}$ AL in the Pantheon data. 

Notice however, that the 30 extrema AL directions of the isotropic Pantheon simulated data are not distributed uniformly. This is due to the fact that the Pantheon SnIa are not isotropically distributed in the sky. As shown in Fig. \ref{fig:distribfigreal} (right panel) the southern (lower) right quadrisphere is almost empty of SnIa datapoints while most of the Pantheon SnIa directions are concentrated in the southern left quadrisphere. This strongly anisotropic distribution of datapoints forces most of the extrema AL directions to concentrate in the southern left - northern right quadrisphere. A possible solution to this problem could the smoothed residual method \cite{Colin:2010ds,Feindt:2013pma,Appleby:2013ida,Appleby:2014lra} that seems to be advantageous in some cases with anisotropically distributed data. This method attempts to ameliorate any anisotropy of the data using a 2D smoothing interpolation of the data on the surface of a unit sphere.

An alternative method to this smoothing approach is to select a more isotropic subset of the full dataset which will be less biased in the selection of the maximum AL direction. Thus, we randomly select a subsample of the Pantheon dataset consisting of 375 SnIa distributed more isotropically in the four quadrispheres (100 in the first three and 75 in the down right quadrisphere) and generate a new reduced dataset (right panel of Fig. \ref{fig:distribfigredreal}). Using this reduced dataset, which is significantly more homogeneous than the full dataset, we produce 100 simulated Pantheon isotropic subsamples using 1500 random directions to split it in two hemispheres and identify the corresponding maximum $\Delta {\cal{M}}/\bar{{\cal{M}}}$ AL magnitudes\footnote{The number of the random directions considered for the identification of the direction of the maximum AL is smaller in this case, since the new dataset is significantly smaller than the original.}. This is illustrated in the left panel of Fig. \ref{fig:distribfigredreal} where we show two opposite points for each maximum AL direction. Clearly, the preferred range of directions disappears completely for the more isotropic subset of the full dataset. However, even in this case where the data are more uniformly distributed, no signal of anisotropy is found, since 33 (red dots in the left panel of Fig. \ref{fig:distribfigredreal}) out of the 100 simulated datasets have larger maximum AL magnitudes of $\Delta {\cal{M}}/\bar{{\cal{M}}}$ than the corresponding magnitude of the real Pantheon data.

The lack of anisotropy signal persists also if we divide the Pantheon data in four redshift bins. Using the same method as described above we construct for each bin 30 isotropic simulated ``Pantheon" datasets and for each dataset we use 1000 random directions to split the sky in two hemispheres and identify the corresponding maximum $\Delta {\cal{M}}/\bar{{\cal{M}}}$ magnitudes. Then we compare the maximum magnitudes $\Delta {\cal{M}}/\bar{{\cal{M}}}$ of the simulated ``Pantheon" datasets with the corresponding maximum magnitude $\Delta {\cal{M}}/\bar{{\cal{M}}}$ of the real data for each bin. The results for each bin are presented in the following Table \ref{tab:hcbinresults}.

\begin{table}[h!]
\caption{The results of the HC method for each bin}
\label{tab:hcbinresults}
\begin{centering}
\begin{tabular}{|c|c|c|}
 \hline 
 \rule{0pt}{3ex}  
Bin  & Redshift Range & Number of Simulated Datasets \\
& & with $\left|\frac{\Delta {\cal{M}}}{\bar{{\cal{M}}}} \right|_{sim}>\left|\frac{\Delta {\cal{M}}}{\bar{{\cal{M}}}} \right|_{real}$\\
    \hline
    \rule{0pt}{3ex}  
$1st$ & $0.01<z<0.13$ & $21/30$\\
$2nd$ & $0.13<z<0.25$ & $8/30$\\
$3rd$ & $0.25<z<0.42$ & $14/30$\\
$4th$ & $0.42<z<2.26$ & $3/30$\\
\hline
\end{tabular}
\end{centering}
\end{table}

\noindent Interestingly, the strongest evidence for anisotropy is not found in the lowest $z$ bin but in the highest $z$ bin $(0.42<z<2.3)$, where only three out of the thirty simulated datasets have larger $\Delta {\cal{M}}/\bar{{\cal{M}}}$ magnitudes than the corresponding magnitude of the real data. However, this mild effect is not statistically significant, since it remains below the $2\sigma$ level.

Even though we have found no evidence for anisotropy of $\Delta {\cal{M}}/\bar{{\cal{M}}}$  in the Pantheon data, the local underdensity scenario as an explanation for the reduced by about $4\%$ low $z$ value of ${\cal{M}}$ pointed out in the previous section, remains viable especially if  we are located close to the center of the underdensity.
Using Eq. \eqref{eq:deltahoverh}, we can also constrain the density contrast $\delta \rho_0/\rho_0$ as well as the dimensionless mater density contrast $\delta \Omega_0/\Omega_0$, through the following coupled system of equations applicable in the context of a LTB model with a cosmological constant assuming a top hat density profile for the void (see Appendix of Ref. \cite{Lukovic:2019ryg})
\begin{align}
&\frac{\delta H_0}{H_0}=\frac{\delta \rho_0}{\rho_0}\left[-0.171-0.322 (\Omega_{0m}-0.3)+0.249(\Omega_{0m}-0.3)^2\right] \nonumber \\
&+\left(\frac{\delta \rho_0}{\rho_0} \right)^2\left[0.031+0.063 (\Omega_{0m}-0.3)\right] -0.022\left(\frac{\delta \rho_0}{\rho_0} \right)^3 \label{eq:deltahoverhwithx} \\ 
&\frac{\delta \Omega_0}{\Omega_0}=\frac{\delta \rho_0}{\rho_0}\left[1.342+0.643 (\Omega_{0m}-0.3)-0.499(\Omega_{0m}-0.3)^2\right] \nonumber \\
&+\left(\frac{\delta \rho_0}{\rho_0}\right)^2 \left[0.367+0.847 (\Omega_{0m}-0.3)\right]+0.056 \left(\frac{\delta \rho_0}{\rho_0}\right)^3  \label{eq:deltaomoveromwithx}
\end{align}
where in this case we set $\Omega_{0m}=0.3153$, \ie the CMB value indicated by the Planck mission \cite{Aghanim:2018eyx}. Substituting $\delta H_0/H_0 \approx 0.02$ [as indicated from Eq. \eqref{eq:deltahoverh}] in Eq. \eqref{eq:deltahoverhwithx}, we calculate $\delta \rho_0/\rho_0=-0.10 \pm 0.04$ and using this value to Eq. \eqref{eq:deltaomoveromwithx} we derive $\delta \Omega_0/\Omega_0=-0.12 \pm 0.02$, in aggreement with previous studies \cite{Lukovic:2019ryg}.

\subsection{Dipole Fitting (DF) Method}
In most physical mechanisms the predicted cosmological anisotropy can be described by a dipole proportional to $cos \theta$. In this case the Dipole Fitting (DF) method \cite{Mariano:2012wx,Chang:2014nca,Lin:2015rza,Deng:2018jrp} is expected to be more sensitive for the detection of the cosmic anisotropy. In this context we define the deviation of the apparent magnitude from its best fit \lcdm values $\bar{m}(z)$ as
\be
\left(\frac{\Delta m (z)}{\bar{m}(z)} \right)_{obs} \equiv \frac{\bar{m}(z)-m(z)}{\bar{m}(z)} \label{eq:appamagrat}
\ee
The basic steps of the DF method are the following \cite{Mariano:2012wx}
\begin{itemize}
\item Convert the coordinates of the SnIa to galactic coordinates $(l,b)$ (they are provided in equatorial coordinates) and define the unit vector $\hat{n}_i$ as
\be 
\hat{n}_i=cos(b_i) \, cos(l_i) \hat{x}+cos(b_i) \, sin(l_i) \hat{y}+sin(b_i) \hat{z}
\ee
\item Define the dipole axis $\vec{D}$ in terms of the parameters $c_1$, $c_2$ and $c_3$ in cartesian coordinates as
\be 
\vec{D}=c_1 \hat{x}+ c_2 \hat{y}+c_3 \hat{z}
\ee
and define
\be 
\left(\frac{\Delta m}{m}\right)_{th}=A \, cos \theta +B \label{eq:dipfitanis}
\ee
where $A$ and $B$ correspond to the dipole and monopole terms of the parametrized anisotropy. The angle $\theta$ is the angle between the datapoint direction of the SnIa with the vector $\vec{D}$ which obeys the relation  
\be 
\hat{n}_i \, \vec{D}=A \, cos \theta_i
\ee
\item Using the maximum likelihood method construct  $\chi^2$ as
\be
\chi^2=\sum_{i=1}^{1048} V^{i} C_{ij}^{-1} V^j 
\ee
where $V^i \equiv \left(\Delta m/m\right)_{obs}-\left(\Delta m/m\right)_{th}=\left[\bar{m}(z_i)-m(z_i)\right]/\bar{m}(z_i)-A \, cos \theta_i-B$ and $C_{ij}$ is the covariance matrix. From the minimization of $\chi^2$, find the best fit values as well as the $1\sigma$ errors of the $c_i$, the monopole term $B$ and thus the dipole term that is derived as $A=\sqrt{\sum_{j=1}^{3}c_j^2}$.
\end{itemize}
We present the results of the application of the DF method for the Pantheon dataset in the following Table \ref{tab:dfresults}.

\begin{table}[h!]
\caption{The best fit values with the $1 \sigma$ error of the $c_i$'s, $A$ and $B$ parameters.}
\label{tab:dfresults}
\begin{centering}
\begin{tabular}{|c|c|}
 \hline 
 \rule{0pt}{3ex}  
Quantity  & Best Fit Value $\pm$ $1 \sigma$ Error\\
    \hline
    \rule{0pt}{3ex}  
$c_1$ & $(-1.41 \pm 3.76) \times 10^{-4}$\\
$c_2$ & $(-0.82 \pm 4.54) \times 10^{-4}$\\
$c_3$ & $(5.28 \pm 7.14) \times 10^{-4}$\\
$A$ & $(5.53 \pm 6.04) \times 10^{-4}$\\
$B$ & $(-0.59 \pm 3.01) \times 10^{-4}$\\
$l$ & $210.254^ \circ \pm 136.564$\\
$b$ & $72.852^ \circ \pm 60.631^\circ$\\
\hline
\end{tabular}
\end{centering}
\end{table}

\noindent The monopole $A$ and dipole $B$ terms are consistent with zero at the $1\sigma$ level. Using the best fit values of the parameters $A$ and $B$, we find the anisotropy direction to be $(l,b)=(210.254^ \circ \pm 136.564, 72.852^ \circ \pm 60.631^\circ)$. Clearly, the errors of the $(l,b)$ coordinates are quite large, covering almost the entire sky area (a result that is in agreement with previous studies \cite{Deng:2018jrp,Zhao:2019azy,Chang:2019utc} using $\Omega_{0m}$ instead of $\cal{M}$).

\begin{figure}[h!]
\centering
\includegraphics[width = 0.45\textwidth]{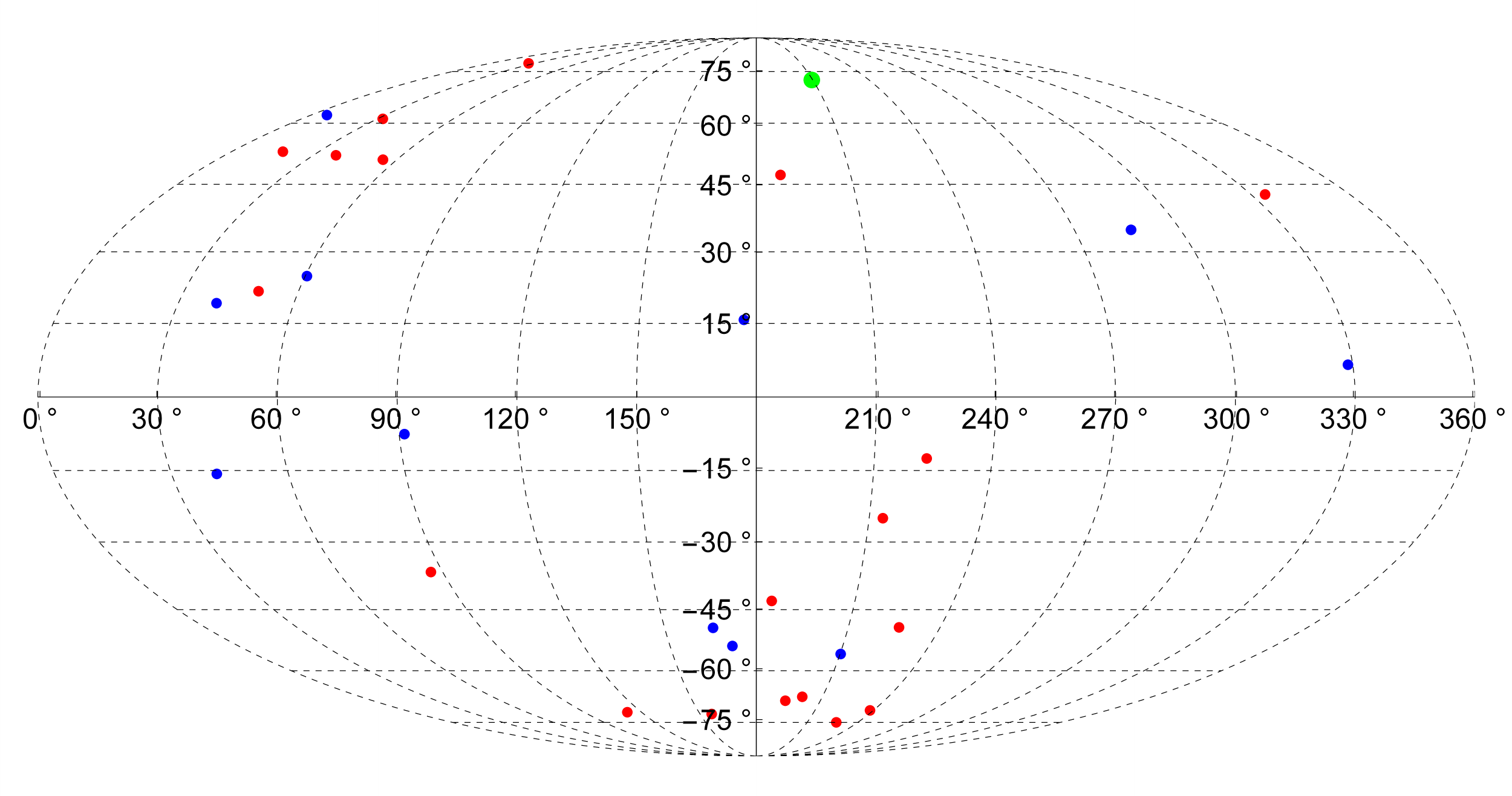}
\caption{The different maximum AL directions corresponding to the 30 random simulated datasets. The green dot corresponds to dipole of the real data, while the blue (red) dots describe the dipole direction of simulated datasets which have smaller (larger) magnitudes of $A$ than the real data. The $1 \sigma$ errors of the $(l,b)$ galactic coordinates are quite large covering almost the entire sky area.}
\label{fig:dfrandomfig}
\end{figure}

The consistency of the derived dipole and monopole terms with statistical isotropy may be investigated using isotropic simulated Pantheon datasets as we did in the context of the HC method. We construct 30 simulated Pantheon datasets as described in the previous subsection, we identify the corresponding dipole anisotropy directions and the best fit values of the parameters $A$ and $B$ as shown in Fig. \ref{fig:dfrandomfig}.

From Fig. \ref{fig:dfrandomfig}, it is clear that no preferred direction is identified since 19 (red points in Fig. \ref{fig:dfrandomfig}) of the 30 isotropic simulated Pantheon datasets have larger dipole magnitudes than the real data. Therefore, we conclude that no statistically significant anisotropy is found using the DF method, in agreement with the corresponding result of the HC method.

\subsection{Comparison of the Two Methods}
From the implementation of the two methods the following useful conclusions, can be extracted 
\begin{itemize}
\item The HC method is more general, since it can detect any kind of anisotropy. On the contrary, the DF method is sensitive only to an anisotropy of the form of Eq. \eqref{eq:dipfitanis}, \ie an anisotropy that has a dipole form.
\item The $1 \sigma$ errors of the anisotropy direction coordinates obtained in the context of the DF method are quite large and cover the entire sky area. Thus, a dipole anisotropy seems to be significantly  disfavoured by the Pantheon dataset indicating that no dipole signal exists in the data. On the contrary, the HC method gives significantly smaller $1 \sigma$ errors as it is tuned for the detection of a much broader range of signals. Therefore, the HC method seems to be more appropriate in order to identify a preferred direction as well as any general anisotropies hidden in the Pantheon data, unless these anisotropies are of the particular dipole form.
\end{itemize}
In conclusion, we have found no evidence of anisotropy in the Pantheon data, a result consistent with previous studies \cite{Saadeh:2016sak,Sun:2018cha,Andrade:2018eta,Deng:2018jrp,Zhao:2019azy,Chang:2019utc}. We have shown however, that the HC method is more appropriate in detecting a general form of anisotropy hidden in the data. We have also demonstrated that the anisotropic distribution of the Pantheon SnIa data leads to a preferred range of anisotropy directions which are detected by the HC method in the context of isotropic simulated Pantheon datasets. This lack of anisotropy does not favour (but also does not exclude) the local underdensity scenario as a possible explanation of the observed reduced value of $\cal{M}$ at low $z$ indicated in section \ref{sec:maxlikelihood}. We thus proceed to examine the alternative mechanism that could lead to a reduced value of $\cal{M}$ at low $z$: the evolving $\mu$ scenario.
\section{Modified Theory of Gravity Scenario}
\label{sec:modgrav}
\begin{figure}[!h]
\centering
\includegraphics[width = 0.47\textwidth]{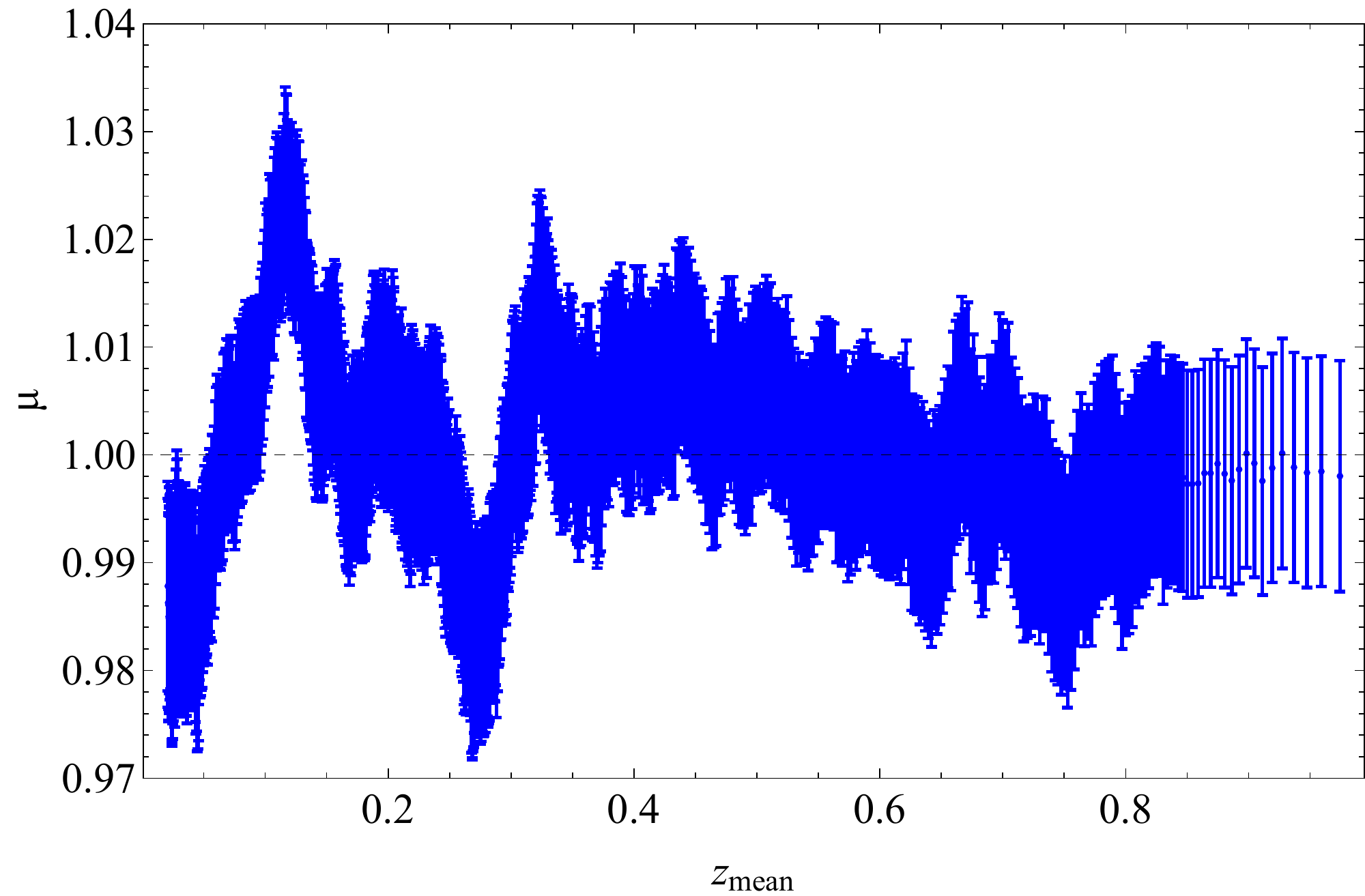}
\caption{The evolution of $\mu$ along with its $1\sigma$ error vs the mean redshift $z_{mean}$ of each 100 datapoints subsample. In the context a modified theory, we detect a $2-3\%$ deviation from the GR predicted value $\mu=1$ (dashed line) at a level up to about $2\sigma$.}
\label{fig:Geffsubsampplot}
\end{figure}

In order to identify the possible evolution of $\mu$ we consider Eq. \eqref{appmagthmodgen} and use the 100 point moving subsample method described in Section \ref{sec:maxlikelihood} (Figure \ref{fig:subsamp}). In particular, we find the best fit value of $M$, fixing $M_0$ to the best fit value of the absolute magnitude $M$ indicated by the full dataset with $h=0.74$. Then, for $b=-3/2$ , we use the best fit values of $M$ and find the corresponding best fit values of $\mu$, thus assigning any redshift dependence of $M$ into a redshift dependence of $\mu$. The resulting best fit values of $\mu$ for each subsample along with the $1\sigma$ errors are shown in Fig. \ref{fig:Geffsubsampplot}.

\noindent Clearly the oscillating behaviour of M at low $z$ shown in Fig. \ref{fig:subsamp} middle panel, is reflected on a corresponding oscillating behaviour for $\mu$ at low $z$ with $\mu<1$ at $z=0$.

\begin{figure}[h]
\centering
\includegraphics[width = 0.47\textwidth]{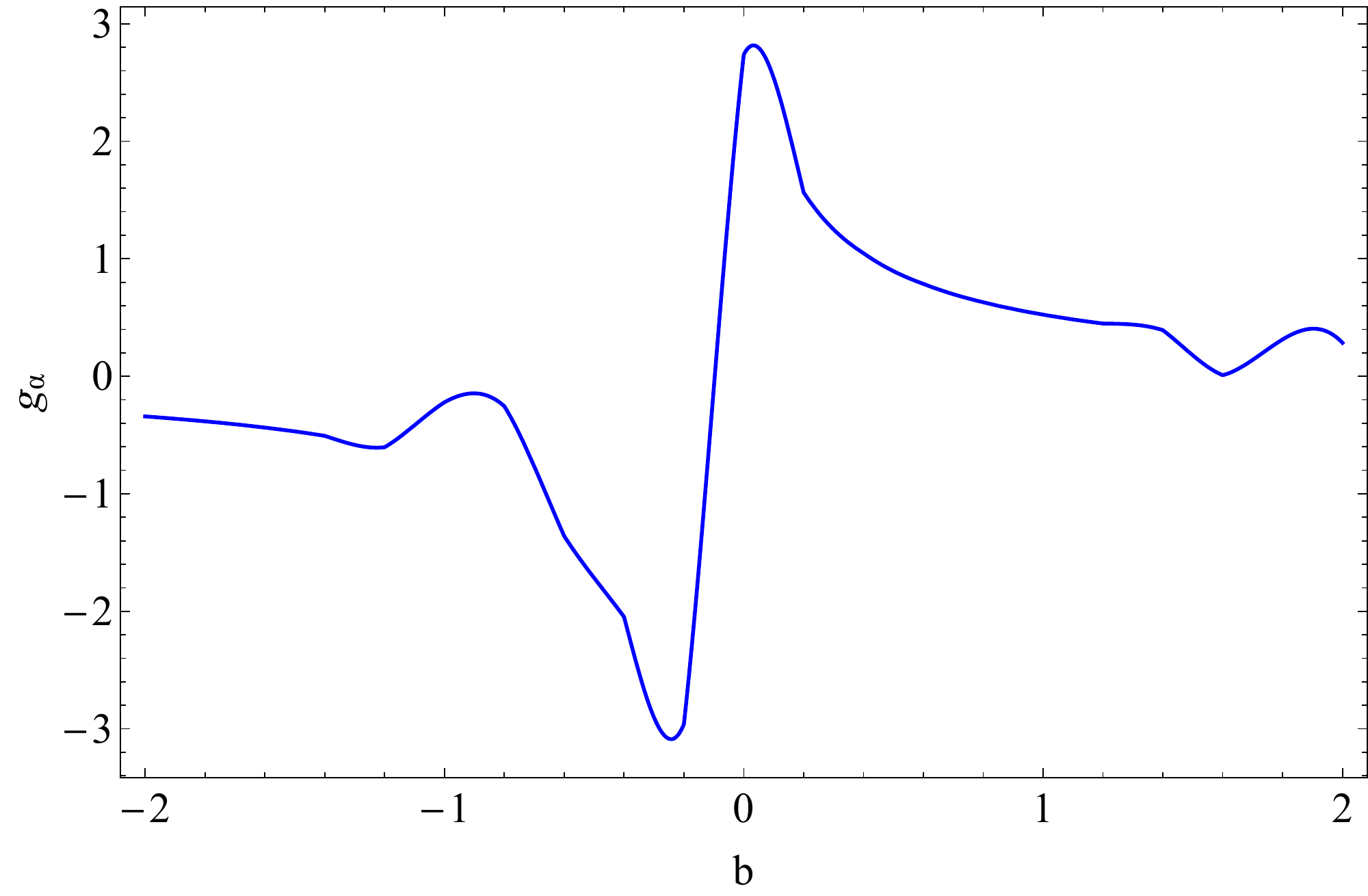}
\caption{The extra parameter $g_a$ as a function of $b$. For $b>0$ we obtain $g_a>0$ while for $b<0$ we find a best fit $g_a<0$.}
\label{fig:gabfig}
\end{figure}

We now consider a $\mu$  parametrization which interpolates GR at early and late times and takes into consideration the solar system and nucleosynthesis constraints. This parametrization is the following \cite{Nesseris:2017vor,Kazantzidis:2018rnb}
\be 
\mu (z,g_a)= 1+g_a \left(\frac{z}{1+z} \right)^2 - g_a \left(\frac{z}{1+z} \right)^4 \label{geffparametrization}
\ee
where $g_a$ is an extra parameter and $z$ is the redshift. Using the modified apparent magnitude \eqref{appmagthmodgen} along with the parametrizaion \eqref{geffparametrization} we construct the corresponding $\chi^2$ function and applying the maximum likelihood method, we obtain the best fit values for the parameters ${\cal{M}},\Omega_{0m},g_a$ and $b$. In this case, $\chi^2$ depends on the same parameters as before $({\cal{M}},\Omega_{0m})$ as well as the extra parameters $g_a$ and $b$. During the minimization we allow $b$ to take various values in the range $-2<b<2$ and interpolate the best fits of the extra parameter $g_a$ as a function of $b$. This is demonstrated in Fig.  \ref{fig:gabfig}.  

\begin{figure*}
\centering
\includegraphics[width = 1.\textwidth]{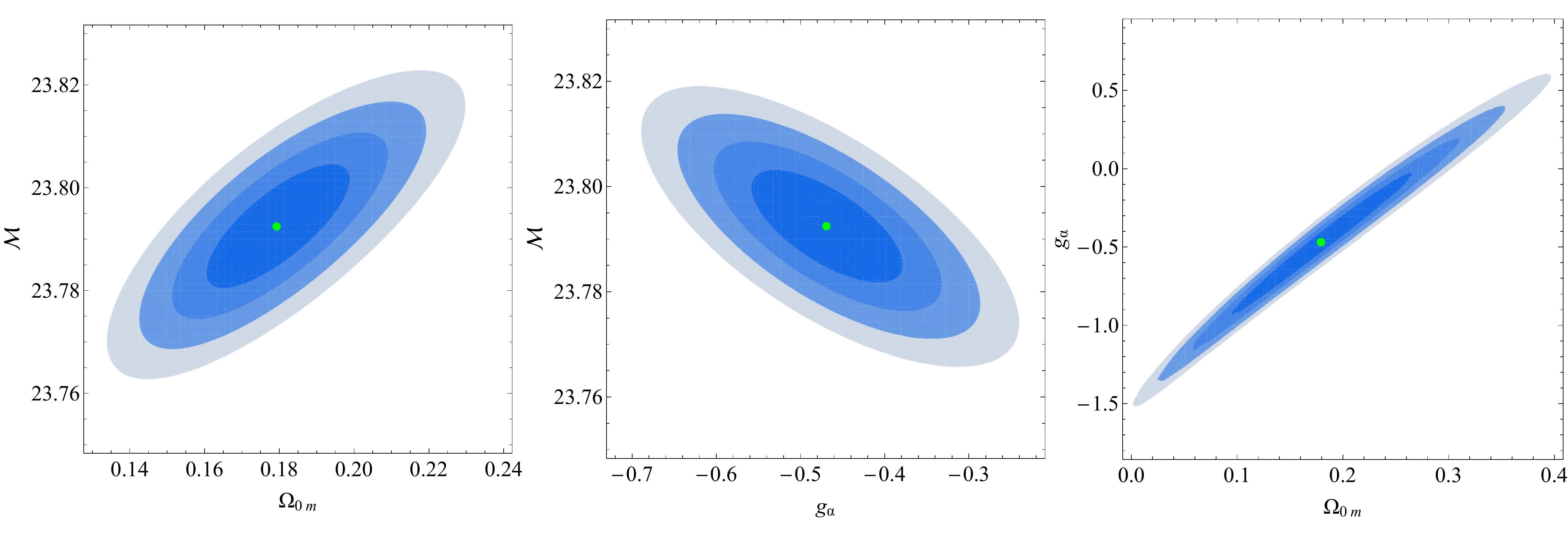}
\caption{The 2D projections of the $1\sigma-4 \sigma$ contours  in the parametric space $({\cal{M}},\Omega_{0m},g_a)$. The projections go through the best fit point (green point) in the 3D parameter space. Notice that the GR point corresponding to $g_a=0$ appears to be more than $4\sigma$ away from the best fit which corresponds to weaker gravity $(g_a<0)$. However, this is a projection effect since in the context of the full 3D parameter space we have $g_a=-0.47 \pm 0.36$.}
\label{fig:contourgeffig}
\end{figure*} 

For negative values of $b$ we obtain negative best fit values for $g_a$ indicating that $\mu<1$, \ie a growth rate that is weaker than expexted in the context of \lcdmnospace.  This result is in agreement with other studies \cite{Nesseris:2017vor,Kazantzidis:2018rnb,Perivolaropoulos:2019vkb,Skara:2019usd} which fit weak lensing and growth data allowing for an evolving $\mu$ and favour an evolving $\mu$ with $g_a<0$.

Setting $b=-3/2$ as indicated by most relevant studies \cite{GarciaBerro:1999bq,Gaztanaga:2001fh} it is straightforward to construct the likelihood parameter contours in the 3D parametric space $({\cal{M}},\Omega_{0m},g_a)$ where ${\cal{M}}$ is given by Eq. \eqref{calmdef}. The best fit parameter values thus obtained are ${\cal{M}}=23.793 \pm 0.009,\Omega_{0m}=0.179 \pm 0.078$ and $g_a=-0.47 \pm 0.36$. In Fig. \ref{fig:contourgeffig} we show the 2D projections of the $1\sigma-4\sigma$ contours  in the parametric space $({\cal{M}},\Omega_{0m},g_a)$. The projections go through the best fit point in the 3D parameter space. Notice that the GR point corresponding to $g_a=0$ appears to be more than $4\sigma$ away from the best fit which corresponds to weaker gravity $(g_a<0)$ in accordance with weak lensing and growth cosmological data  \cite{Hildebrandt:2016iqg,Joudaki:2017zdt,Kohlinger:2017sxk,Abbott:2017wau,Nesseris:2017vor,Kazantzidis:2018rnb,Perivolaropoulos:2019vkb,Skara:2019usd}. However, this is a projection effect since in the context of the full 3D parameter space we have $g_a=-0.47 \pm 0.36$, a value that is approximately $1.5 \sigma$ away from the GR predicted one.

\section{Conclusion - Discussion - Outlook}
\label{sec:concl}
We have performed a redshift tomographic analysis of the latest SnIa (Pantheon) data in the context of a \lcdm model fitting simultaneously  the matter density parameter $\Omega_{0m}$ and  the parameter $\cal{M}$ which depends on both the calibrated absolute magnitude $M$ and the Hubble parameter $H_0$. Including only statistical uncertainties of the Pantheon data, we have found a mild tension $(2-3 \sigma)$  between  the best fit value of $\cal{M}$ obtained from low $z$ SnIa $(z \in [0.01,0.2])$ and the corresponding value obtained from the full Pantheon dataset. This  deviation drops to slightly more than $1\sigma$ when the systematic uncertainties are taken into account (see the Appendix \ref{sec:Appendix_A}). If this mild tension is not a statistical fluctuation it could be either explained as a locally higher value of $H_0$ corresponding to a local underdensity with $\left(\frac{\delta \rho_0}{\rho_0}, \frac{\delta \Omega_0}{\Omega_0}\right) \simeq \left(-0.10 \pm 0.04, -0.12 \pm 0.02\right)$ or as a modified gravity effect leading to a time variation of  Newton's constant. 

In the context of the local underdensity scenario, a degree of anisotropy is anticipated for $\cal{M}$, unless the observer is located at the center of this underdensity. Thus we used two methods to search for statistically significant anisotropy in the Pantheon SnIa data: The Hemisphere Comparison (HC) method and the Dipole Fit (DF) method. Even though we found no statistically significant evidence for cosmological anisotropy our analysis revealed the following interesting facts:
\begin{itemize}
\item Using simulated Pantheon-like data constructed under the assumption of an underlying \lcdm model, we showed that the anisotropic distribution of the SnIa datapoints in the sky generically favours the range anisotropy of directions consistent with the dataset in $b\in \left[-15 ^\circ, -45 ^\circ\right]$, $l\in \left[60 ^\circ,150^\circ\right]$ (or in the opposite direction $b\in \left[15 ^\circ, 45 ^\circ\right]$, $l\in \left[240 ^\circ,330^\circ\right]$). We constructed a more isotropically distributed subset of the Pantheon data that appears to be free of this limitation but less powerful in detecting overall anisotropy due to reduced number of datapoints.
\item The HC method appears to be more powerful in detecting a general anisotropy signal than the DF method since the statistical uncertainties for both the magnitude and direction of anisotropy appear to be smaller in the context of this method.
\end{itemize}
The lack of evidence for statistical anisotropy in the SnIa data does not favour the local underdensity scenario as a possible explanation for the reduced value of $\cal{M}$ at low $z$ as such anisotropy would be expected in the context of a non-spherical anisotropy and/or an off-center observer.

The abnormal variation of $\cal{M}$ can also be explained in the context of a modified theory of gravity with an evolving Newton's constant at low redshifts. In this context, allowing for an evolving normalized Newton's constant $\mu(z)$, we found a $2\sigma$ deviation of $ \sim 2-3 \%$ from the GR predicted value at low redshifts. Moreover, considering a physically motivated parametrization for the evolving Newton's constant that interpolates GR at early and late times [Eq. \eqref{geffparametrization}], we derived the best fit value of the extra parameter $g_a$ as $g_a=-0.47 \pm 0.36$. This value is approximately $1.5 \sigma$ away from the GR predicted value ($g_a=0$), favouring a reduced Newton's constant compared to GR. This weak hint is consistent with the results at low $z$ of other studies that  mildly favour weakening gravity using growth \cite{Nesseris:2017vor,Kazantzidis:2018rnb,Kazantzidis:2019dvk,Skara:2019usd} and weak lensing data \cite{Hildebrandt:2016iqg,Joudaki:2017zdt,Kohlinger:2017sxk,Abbott:2017wau}.

\begin{figure*}
\centering
\includegraphics[width = 1.\textwidth]{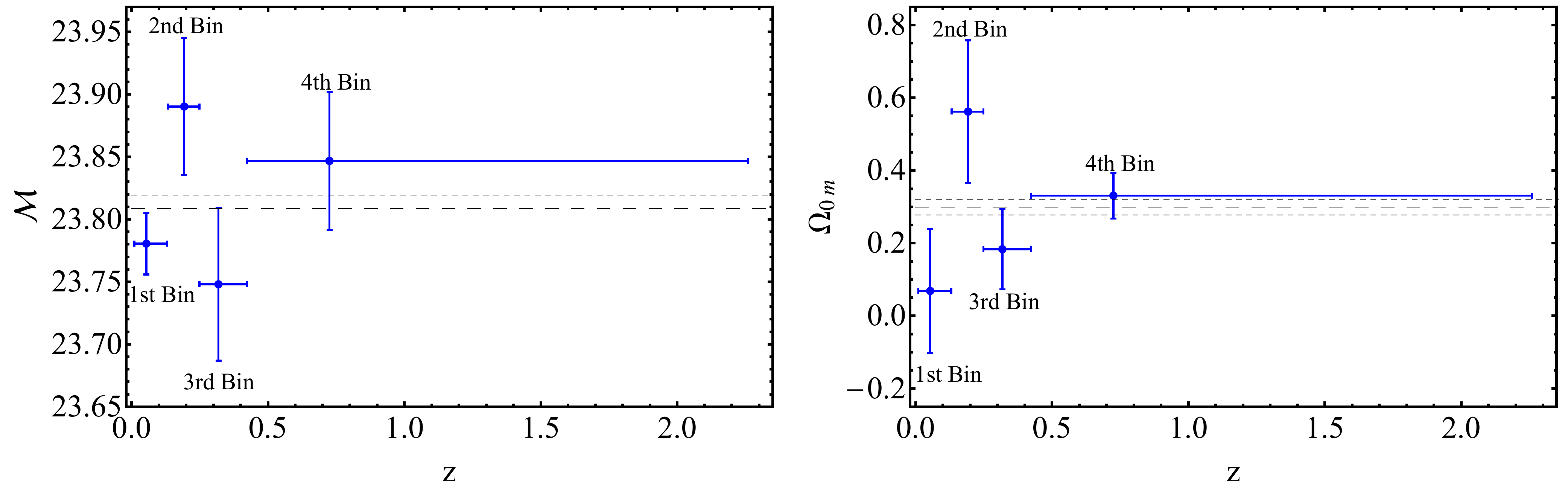}
\caption{The best fit values of $\cal{M}$ (left panel) and $\Omega_{0m}$ (right panel) as well as the $1 \sigma$ errors for the four bins, including the systematic uncertainties. Notice that the oscillating behaviour remains and it is still highly improbable in the context of constant underlying $\cal{M}$ and $\Omega_{0m}$.}
\label{fig:crossplot_2}
\end{figure*}

Interesting extensions of the present analysis include the following:
\begin{itemize}
\item
Use of an extended up to date SnIa dataset with more uniform distribution in the sky to investigate and further constrain the possible evolution of the parameter $\cal{M}$ with redshift and its connection with a possible cosmic anisotropy.
\item 
Further investigate the connection between an evolving $\cal{M}$ and an evolving Newton's constant in the context of various models for the mechanism of SnIa explosion. In particular a reliable estimate of the sign and value of the power index $b$ that connects the evolving absolute magnitude $M$ with the effective Newton's constant is important for imposing reliable constraints on modified gravity models from the possible evolution of SnIa absolute luminosity.
\item
The use of alternative standard candle probes (\eg $\gamma$ ray bursts) to search for possible similar hints of variation of $H_0$ and/or $\Geff$. 
\item
The identification of new statistical tests probing for cosmological anisotropies of SnIa data and the comparison of their efficiency with the standard methods used in the present analysis (HC and DF).
\item The consideration of alternative background expansion cosmologies. It may be possible to absorb the variation of $\cal{M}$ at low $z$ in the context of a varying dark energy equation of state parameter $w$ at low $z$. In this context, a varying $\cal{M}$ at low $z$ may be a hint for a variation of $w$. Such variation may also play a role in resolving the $H_0$ problem \cite{Yang:2018qmz,Li:2020ybr,Benevento:2020fev}.
\end{itemize}

\textbf{Numerical Analysis Files}: The numerical files for the reproduction of the figures can be found in \cite{numcodes}.

\section*{Acknowledgements}
We thank Savvas Nesseris for useful discussions.  This research is co-financed by Greece and the European Union (European Social Fund- ESF) through the Operational Programme ``Human Resources Development,
Education and Lifelong Learning" in the context of the project ``Strengthening Human Resources Research Potential via Doctorate Research – 2nd Cycle" (MIS-5000432), implemented by the State Scholarships Foundation (IKY) and through the Operational Programme ``Human Resources Development, Education and Lifelong Learning 2014-2020" in the context of the project No. MIS 5047648.

\appendix
\section{Correction with Systematic Uncertainties}
\label{sec:Appendix_A}
Including the systematic uncertainties, the best fit parameters indicated by the full dataset are ${\cal{M}}=23.809 \pm 0.011$ and $\Omega_{0m}=0.299 \pm 0.022$. Clearly, the inclusion of systematic uncertainties increases the best fit parameters as well as the $1\sigma$ errors of the parameters \cite{Scolnic:2017caz}. Thus, we construct Fig. \ref{fig:crossplot_2} which corresponds to Fig. \ref{fig:crossplot} and takes into account the systematic uncertainties.

Clearly, the oscillating trend that was present in Fig. \ref{fig:crossplot} remains.  In the context of a constant underlying $\cal{M}$ and $\Omega_{0m}$ this  large amplitude oscillating behaviour remains a highly unlikely event since all three lowest $z$ bins differ by more than $1 \sigma$ from their expected values for both $\cal{M}$ and $\Omega_{0m}$. 

\raggedleft
\bibliography{Bibliography}

\end{document}